\documentclass[journal]{IEEEtran}
\usepackage{amsmath}
\usepackage{amsfonts}
\usepackage{amssymb}
\usepackage{mathrsfs}
\usepackage{makecell,multirow,diagbox}
\usepackage{flushend}
\usepackage{color}

\newcounter{RomanNumber}

\usepackage{cite}

\ifCLASSINFOpdf
   \usepackage[pdftex]{graphicx}
\else
\fi
\usepackage{amsmath}
\usepackage{algorithmicx,algorithm}
\usepackage{algpseudocode}
\ifCLASSOPTIONcompsoc
\usepackage[caption=false,font=normalsize,labelfont=sf,textfont=sf]{subfig}
\else
  \usepackage[caption=false,font=footnotesize]{subfig}
\fi
\begin{document}
\title{A Time Efficient Approach for Decision-Making Style Recognition in Lane-Change Behavior}
%
%
%

\author{Sen Yang, Wenshuo Wang,~\IEEEmembership{Member, IEEE}, Chao Lu, Jianwei Gong~\IEEEmembership{Member, IEEE}, and Junqiang Xi 
\thanks{The first two author, S. Yang and W. Wang, contribute equally to this work. This work is supported by the National Natural Science Foundation of China (No. 91420203, 61703041). ({\it Corresponding Author: Junqiang Xi})}
\thanks{S. Yang, J. Gong, and J. Xi are with the Department of Mechanical Engineering, Beijing Institute of Technology, Beijing, China, 100081. email:yangsen1990@bit.edu.cn; xijunqiang@bit.edu.cn.}
\thanks{W. Wang is with the Department of Mechanical Engineering, Carnegie Mellon University, Pittsburgh, PA, 15232 USA. email:wwsbit@gmail.com.}
\thanks{C. Lu is with the Department of Mechanical Engineering, Beijing Institute of Technology, Beijing, China 100081, and also with the Advanced Vehicle Engineering Centre, Cranfield University, Cranfield MK43 0AL, U.K. email:chaolu@bit.edu.cn.}
}

\maketitle



\begin{abstract}
Fast recognizing driver's decision-making style of changing lanes plays a pivotal role in safety-oriented and personalized vehicle control system design. This paper presents a time-efficient recognition method by integrating $ k $-means clustering ($ k $-MC) with K-nearest neighbor (KNN), called $ k $MC-KNN. The mathematical morphology is implemented to automatically label the decision-making data into three styles (\text{moderate}, \text{vague}, and \text{aggressive}), while the integration of $ k $MC and KNN helps to improve the recognition speed and accuracy. Our developed mathematical morphology-based clustering algorithm is then validated by comparing to agglomerative hierarchical clustering. Experimental results demonstrate that the developed $ k $MC-KNN method, in comparison to the traditional KNN, can shorten the recognition time by over 72.67\% with recognition accuracy of 90\%$\sim$98\%. In addition, our developed $ k $MC-KNN method also outperforms the support vector machine (SVM) in recognition accuracy and stability. The developed time-efficient recognition approach would have great application potential to the in-vehicle embedded solutions with restricted design specifications.

\end{abstract}

%
\IEEEpeerreviewmaketitle

\section{Introduction}
%
%
%
%
\subsection{Motivation}
\IEEEPARstart{M}{aking}  a human-friendly decision of changing lanes is crucial to intelligent vehicle control\cite{martinez2017driving}, traffic efficiency and road safety\cite{li2017studies}, and human-like autonomous driving systems\cite{nilsson2017lane}. Human driver will generate and carry out various decision-making policies to determine if, when, and how to change lanes \cite{nilsson2016if} by evaluating the current driving situations according to their internal model\cite{zhou2016does}. Modeling such various drivers' decision-making processes is a nontrivial task for applications. It is relatively easy to establish personalized models for a small group of drivers by using advanced learning methodologies\cite{schnelle2017personalizable}, but not feasible for hundreds of millions of human drivers due to excessive cost of time and resources. Building a model for each group of drivers with similar driving characteristics, instead a single model for each driver, will be a cost-effective solution to this issue. In other words, classifying decision-making style into several distinguishable groups empowers us to efficiently describe large amounts of human drivers at low cost. Therefore, it is necessary to classify drivers into groups and then analyze their lane change behavior.

\subsection{Related Research}
In general, a complete lane change task consists of two parts: decision-making and action-execution. Many existing research concerning drivers' operation style has been conducted, for instance, subjectively classifying and labeling drivers' steering signals of double lane-changing maneuvers into several groups according to prior knowledge of driving style\cite{schnelle2017personalizable,wang2017human}. These labeled data were then used to train a classifier based on supervised learning methodologies such as SVM \cite{19} and fuzzy logic \cite{22}.

In terms of decision-making, drivers will prefer different lane-change strategies, depending on their mandatory and discretionary demands \cite{pan2016modeling,keyvan2016categorization} as well as the driving situations\cite{balal2016binary}. Sun \textit{et al}.\cite{Sun2012Lane} conducted an instrumented vehicle-based experiment and found that the urban arterial lane-changing decision-making process heavily depends on driver characteristics. They also proposed a comprehensive framework of modeling drivers' lane-changing maneuvers\cite{Sun2010Research} by computing the lane-changing probability for each scenario over different driver types\cite{Sun2014A}. To enrich the lane change models in traffic simulation packages, Keyvan-Ekbatani, \textit{et al}. \cite{keyvan2016categorization} categorized drivers' lane change strategies into different groups based on a two-stage framework consisting of testing on-line and reviewing off-line, but this method could lead to subjective and empirical results. In addition, a driving style questionnaire was implemented by giving scores to the surveyed questions for distinguishing drivers\cite{23}. The aforementioned classification methods are supervised, but they are commonly time-consuming to manually label huge amounts of driving data. In order to improve classification performance with little labeling efforts, a semi-supervised SVM was developed to classify drivers into aggressive and moderate driving styles with a few labeled data among all collected driving data \cite{20}. However, it is applicationally intractable to prepare objective annotations for training data since this approach does not fully rule out the personal subjective impacts.
 
Numerous logic-based methods have been applied to improve recognition performance but they are computationally expensive and require prior knowledge of data. For example, fuzzy reasoning methods were used to infer driver's lane-change intent\cite{balal2016binary} and identify driving style \cite{22}, which highly relays on the prior knowledge and experience of data analysts and their observations, statistics, and analysis \cite{27}. On the other hand, advanced machine learning techniques were also implemented to recognize driving style. For example, Miro Enev, \textit{et al}. \cite{28} applied four machine learning algorithms, including SVM, random forest, naive Bayes, and KNN, to recognize drivers' driving styles. Wang, \textit{et al}. \cite{1} proposed a pattern-recognition algorithm by combining \emph{k}-MC and SVM together to shorten recognition time and improve recognition performance. In order to deal with the uncertainty of driver behavior in driving style recognition, a statistical-based recognition method using Bayesian probability and kernel density estimation was proposed \cite{Wang2016Statistical}. Zhang, \textit{et al}. \cite{11} investigated three direct pattern-recognition approaches to classify driver's steering operation skills in the double lane-change task, including multilayer perception artificial neural networks, decision tree, and SVM. More state-of-the-art literature related to driving style recognition refers to the literature \cite{martinez2017driving,wang2014modeling,wang2018driving}. These above mentioned algorithms greatly improve the recognition accuracy; however, they usually require a long time to train models to obtain satisfied results \cite{1}, especially when dealing with big data. 

In the above discussed literature, they mainly have two limitations: (1) Requiring sufficient prior knowledge to manually label training data, which is practically intractable for multidimensional large-scale driving data\cite{appenzeller2017scientists9,wang2017much}; (2) Learning classifiers and recognizing driving style for a new observation takes a long time, which impedes these algorithms from being used in online applications.

\subsection{Contributions}
This paper aims to develop a time-efficient way to improve recognition performance of drivers' decision-making style in lane-changing scenarios with little subjective interference involved while labeling training data. Our main contributions cover two aspects, listed as follows.
\begin{enumerate}
	\item Proposing an unsupervised method based on mathematical morphology to label training data, which does not require the prior knowledge of clusters or other parameters, thereby reducing the efforts of tagging data and excluding the subjective influences of data analysts.
	\item Developing a $k$-means clustering-based K-nearest neighbor ($k$MC-KNN) method to accelerate the recognition process and thus shorten recognition time.
\end{enumerate}

\subsection{Paper Organization}
This paper is organized as follows. Section II presents the mathematical morphology method and the $k$MC-KNN method. Section III describes the driving scenarios and data collection. Section IV shows the experimental results. Finally, conclusions are presented in Section V.
\section{Classification and Recognition Methods}
This section will present the approaches to training data auto-classification and new data recognition. First, we will detail the mathematical morphology method which can automatically label the collected data without any prior knowledge. Then we will show our proposed $k$MC-KNN method for driving style recognition.

\subsection{Classification Method}
 The mathematical morphology was primarily constructed as a non-linear processing and analysis tool \cite{serra2012mathematical} for image segmentation in order to obtain a good description and representation of the shapes of these segments\cite{management2018ophthalmology}. Its expanding applications also cover, for instance, boundary detection, automatic image segmentation and reconstruction, pattern recognition, and signal and image decomposition\cite{33}. Inspired by its advantages and these applications, in this paper, we employ the fundamental operators of mathematical morphology\cite{34} -- \textit{dilation} process and \textit{erosion} process -- to search data with the same characteristics and then cluster them.
 
\subsubsection{Dilation and Erosion}
Given an original set $ \textbf{A} \subset \mathbb{Z}^{d} $ and a kernel set $ \textbf{B}(x) =\textbf{B}^{\vee}_{x}=\{b-x:b\in\textbf{B}\} $ with the point $x$ as its origin\cite{34}, the morphological dilation and erosion of \textbf{A} by \textbf{B} are defined as (\ref{eq:1}) and (\ref{eq:2}), respectively.  
\begin{eqnarray}
\textbf{A} \oplus \textbf{B}\equiv \{x:\textbf{B}^{\vee}_{x}\cap \textbf{A}\neq\varnothing\}
\label{eq:1}
\end{eqnarray}
\begin{eqnarray}
\textbf{A} \ominus \textbf{B}\equiv \{x:\textbf{B}^{\vee}_{x}\subseteq \textbf{A}\}
\label{eq:2}
\end{eqnarray}

\begin{figure}[t]
	\centering
	\subfloat[\textbf{A}]{\includegraphics[width=0.10\textwidth]{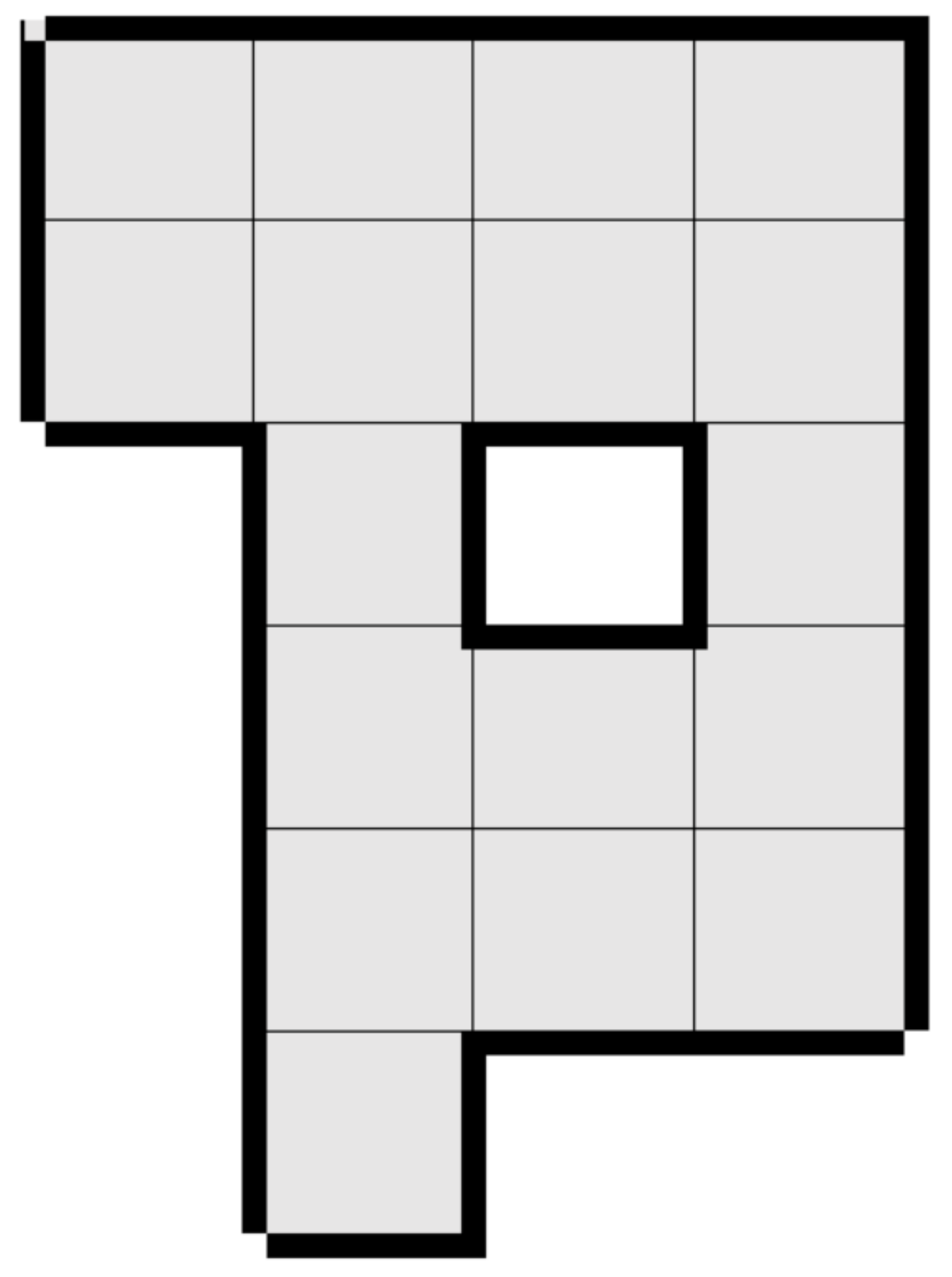}\label{}}
	\hfil
	\subfloat[\textbf{B}]{\includegraphics[width=0.05\textwidth]{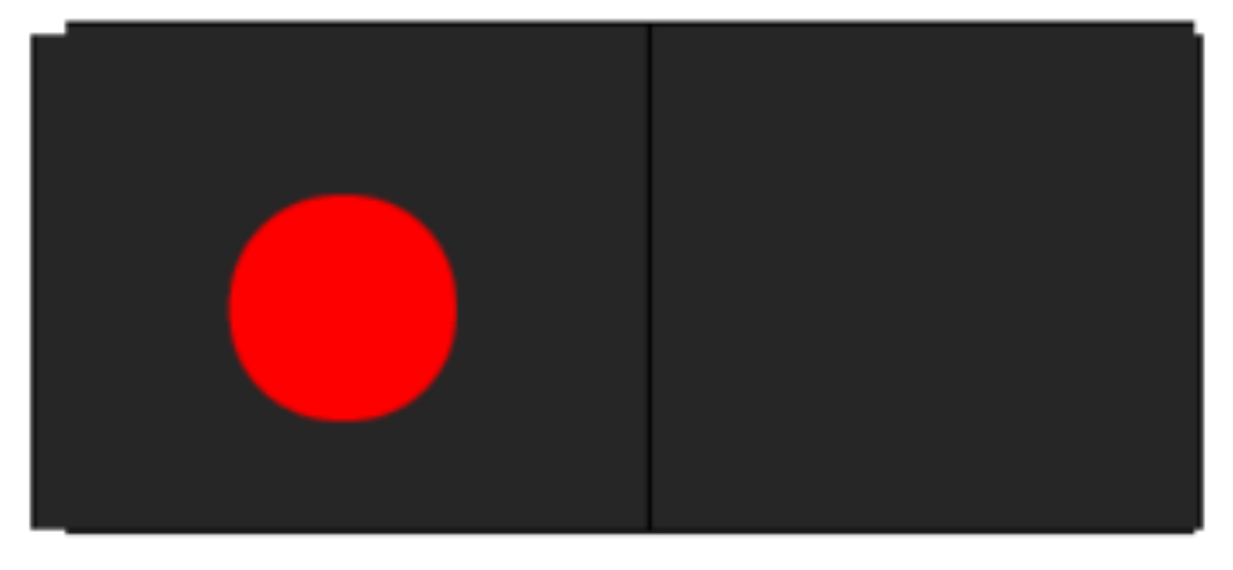}}
	\hfil
	\subfloat[$\textbf{A} \oplus \textbf{B}$]{\includegraphics[width=0.47\textwidth]{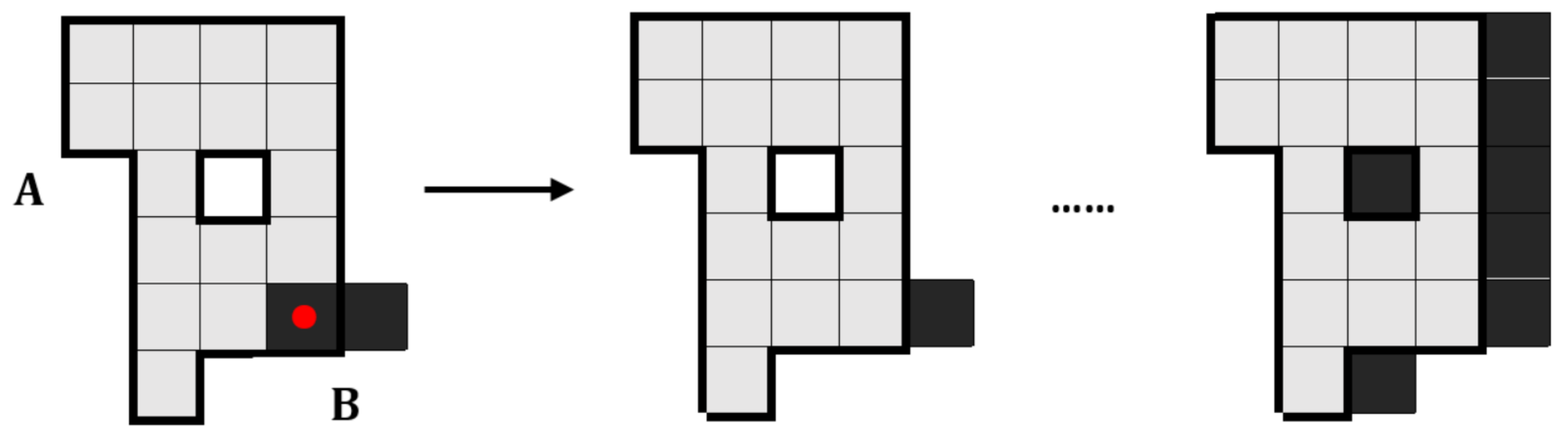}}
	\hfil
	\subfloat[$\textbf{A} \ominus \textbf{B}$]{\includegraphics[width=0.47\textwidth]{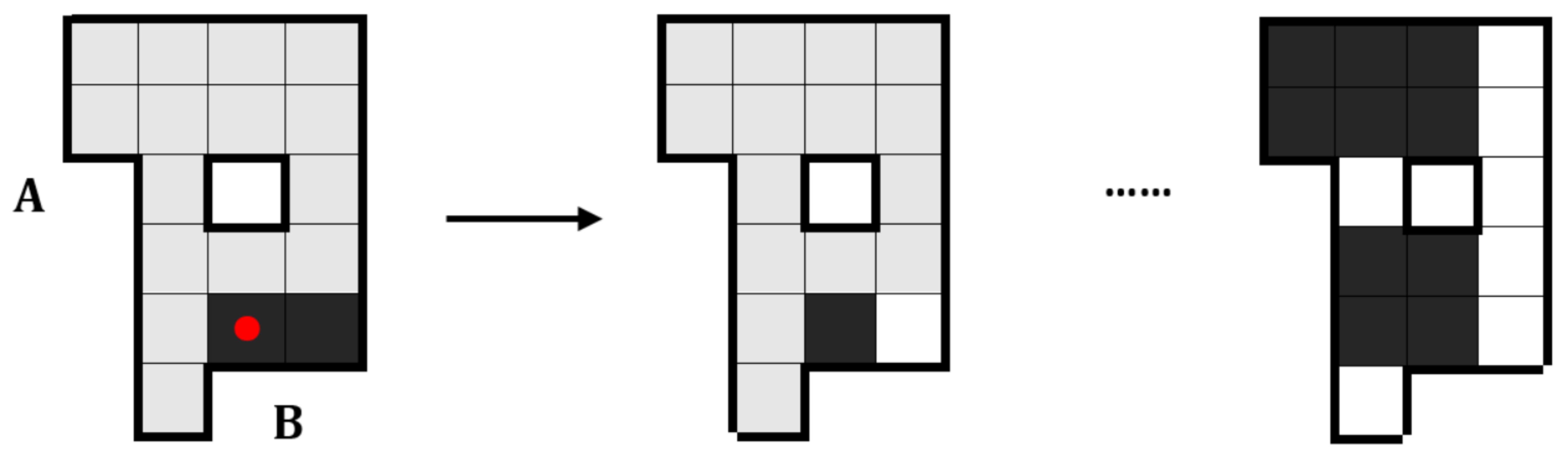}}

	\caption{Illustration of dilation and erosion operations. (a) The original set \textbf{A}. (b) The kernel set \textbf{B}. The pixel marked with a red circle is the origin $x$. (c) The dilation process of \textbf{A} by \textbf{B}. The pixels that were added by the dilation operations are marked in light black. (d) The erosion process of \textbf{A} by \textbf{B}. The pixels retained by the erosion operation are marked in light black.}
	\label{fig.2}
\end{figure}

In order to understand the operations of dilation and erosion, a visualized example is shown in Fig. \ref{fig.2}. The dilation process of \textbf{A} by \textbf{B} (Fig. \ref{fig.2}c) is achieved by adding the pixels of \textbf{B} into \textbf{A} when the origin pixel of \textbf{B} goes through \textbf{A}, while the erosion process of \textbf{A} by \textbf{B} (Fig. \ref{fig.2}d) is achieved by removing these pixels of \textbf{A} which \textbf{B} is not completely contained. From this example, it can be known that the dilation operation merges the points around the target area (\textbf{A}) to fill small holes in the area and small depressions at the edges of the area, while the erosion operation removes these horns smaller than the kernel structure (\textbf{B}).
 
\subsubsection{Mathematical Morphology-Based Clustering Algorithm}
In order to explore objectively irregular clusters of driving style, we develop a clustering algorithm by making full use of dilation and erosion, which can discover such clusters with arbitrary shape and automatically determine the number of clusters by making full use of underlying data information\cite{35,Luo2010Clustering}. Given a data set $ \{\mathbf{x}^{(n)}\}_{n=1}^{N} $, where $N\in \mathbb{N}^{+}$ is the length of data and $\mathbf{x}=(x_1,\dots,x_i, \dots, x_I)$ is a vector with $I\in \mathbb{N}^{+}$ variables, the procedure of the mathematical morphology-based clustering algorithm can be achieved by following steps.

\begin{enumerate}
	\item Given the data set $  \{\mathbf{x}^{(n)}\} $, we get the normalized data set $\{\mathbf{\overline{x}}^{(n)}\}$ by (\ref{eq:3}), and then transform $\{\mathbf{\overline{x}}^{(n)}\}$ into a positive integer set $\{\mathbf{\widetilde{x}}^{(n)}\}$ with value between $1$ and $q+1$ by (\ref{eq:4}).
	\begin{eqnarray}
	\overline{x}_{i}^{(n)}=\dfrac{x_i-x_{\min}}{x_{\max}-x_{\min}}
	\label{eq:3}
	\end{eqnarray}
	\begin{eqnarray}
	\widetilde{x}_{i}^{(n)}=\mathrm{fix} (\overline{x}_{i}^{(n)}\times q_i)+1
	\label{eq:4}
	\end{eqnarray}
	where $n=1,\dots,N, i=1,\dots,I$; $x_{\min}$ and $x_{\max}$ are the minimum and maximum of $x_i$, respectively. $\overline{x}_{i}^{(n)}$, $\widetilde{x}_{i}^{(n)}$ are the elements of $\mathbf{\overline{x}}^{(n)}$ and $\mathbf{\widetilde{x}}^{(n)}$, respectively. $\mathrm{fix}(\cdot)$ in (\ref{eq:4}) is the truncated function, $q_i$ is a suitable integer for the parameter $x_i$.
	
	\item Then, set matrix $\textbf{A}_{q_1\times q_2\times\dots\times q_I}$ with $\textbf{A}(\mathbf{\widetilde{x}}^{(n)} )=1$ if the element of $\textbf{A}$ is equal to $ \mathbf{\widetilde{x}}^{(n)} $ and otherwise $0$. Until now, the given data is converted to the original matrix \textbf{A} filled by 0 or 1. For \textbf{B}, we choose the kernel matrix to be spherical with a suitable radius $r$, which is much smaller than the original matrix \textbf{A}.
	
	\item The dilation result  $\textbf{A}_1$ of $\textbf{A}$ by $\textbf{B}$ is then obtained by (\ref{eq:1}), and $\textbf{A}_2$ is the result of the erosion of $\textbf{A}_1$ by $\textbf{B}^{'}$ whose radius is $r+1$ based on (\ref{eq:2}). These collected areas in $\textbf{A}_2$ are clusters, and the number of collected areas, denoted as $J$, is the number of clusters. The cluster with a small amount of data is regarded as the noise and then removed from $\textbf{A}_2$.
	
	\item The data $\mathbf{x}^{(n)}$ with the shortest Euclidean distance between cluster centers $C_j$ and $\mathbf{x}^{(n)}$ belongs to the cluster $j$, computed by (\ref{eq:5})
	\begin{eqnarray}
	\hat{j}^{(n)}=\arg\min_j \Arrowvert \mathbf{x}^{(n)}-C_j\Arrowvert
	\label{eq:5}
	\end{eqnarray}
	with $ j=1,2,...,J$.
\end{enumerate}

\subsection{Recognition Method}
Before introducing $ k $MC-KNN, we shall present the preliminaries of KNN and $ k $MC.
\subsubsection{KNN}
KNN is a non-parametric classification method, which has been widely used in many research fields such as text categorization and image processing. Given a labeled dataset $\mathcal{M}=\{(\mathbf{x}^{(n)} ,y^{(n)} )\}_{n=1}^{N}$  and a new data $\mathbf{x}^{(*)}$, where $y^{(n)}\in\{y_j\}_{j=1}^{J}$ are the labels of training samples and $\{y_j\}$ is the set of labels, the KNN algorithm is described as follows.

\begin{enumerate}
	\item Normalize the training samples $\mathbf{x}^{(n)}$ and the data $\mathbf{x}^{(*)}$ using (\ref{eq:3}), then attain the normalized training sample set $\{\mathbf{\overline{x}}^{(n)}\}$ and the normalized new data $\mathbf{\overline{x}}^{(*)}$, respectively.
	
	\item Evaluate the similarity levels between the training samples $\mathbf{x}^{(n)}$ and the data $\mathbf{x}^{(*)}$ using (\ref{eq:6}). And then choose $K=\sqrt{N}$ of the most similar samples as the KNN collection of $\mathbf{x}^{(*)}$.
	\begin{eqnarray}
	\mathrm{SIM}(\mathbf{x}^{(n)},\mathbf{x}^{(*)})=\Arrowvert \mathbf{\overline{x}}^{(n)}-\mathbf{\overline{x}}^{(*)}  \Arrowvert^2
	\label{eq:6}
	\end{eqnarray}
	
	\item Decide the category of the given data $\mathbf{x}^{(*)}$ using
	\begin{eqnarray}
	\hat{j}^{(*)}=\arg \max_j\sum_{n=1}^{K}\mathrm{SIM}(\mathbf{x}^{(n)},\mathbf{x}^{(*)})\cdot\rho(\mathbf{x}^{(n)},y_j)
	\label{eq:13}
	\end{eqnarray}
	with
	\begin{eqnarray}
	\rho(\mathbf{x}^{(n)},y_j)=\left\{
	\begin{aligned}
	1,\ \mathrm{if}\ y^{(n)}=y_j\\
	0,\ \mathrm{if}\ y^{(n)}\neq y_j
	\end{aligned}
	\right.
	\nonumber
	\end{eqnarray}
	where $j=1,2,\dots,J$ and $n=1,2,\dots,K$.
\end{enumerate}

The major limitation of KNN is that a large number of design vectors in the trained classifier will significantly increase computational complexity for recognizing new data samples, which impedes its applications to vehicle dynamics wherein high-dimension variables are required for classification. 

\subsubsection{$ k $-MC}
MacQueen \cite{37} first proposed the $ k $-MC algorithm, which partitions the given $n$ objects into $k$ clusters with each object belonging to the cluster with the nearest mean. The $ k $-MC includes four basic steps: initialization, assignment, update, and repeat. Given a data set  $ \{\mathbf{x}^{(n)}\}_{n=1}^{N} $, four steps should be taken as follows:

\begin{itemize}
	\item \textit{Initialization}: Normalize $ \{\mathbf{x}^{(n)}\}$ using (\ref{eq:3}), and get the normalized data $\{\mathbf{\overline{x}}^{(n)}\}_{n=1}^{N} $. Randomly choose $k$ instances $\{\mathbf{m}^{(\eta)}_1\}_{\eta=1}^{k}$ from $\{\mathbf{\overline{x}}^{(n)}\}$ as the initial conditions.
	
	\item \textit{Assignment}: Assign each data point $\mathbf{x}^{(n)}$ to the nearest cluster according to the $\hat{\eta}^{(n)}$.
	\begin{eqnarray}
	\hat{\eta}^{(n)}=\arg\min_{\eta}\Arrowvert \mathbf{m}^{(\eta)}_t-\mathbf{\overline{x}}^{(n)} \Arrowvert
	\label{eq:9}
	\end{eqnarray}
	
	\item \textit{Update}: Adjust the means $\mathbf{m}^{(\eta)}_t$ to match the sample means of the data points through the following formula.
	\begin{eqnarray}
	\mathbf{m}^{(\eta)}_t=\frac{\sum_{n}\beta^{(n)}_{\eta} \mathbf{\overline{x}}^{(n)}}{\sum_{n}\beta^{(n)}_{\eta}}
	\label{eq:10}
	\end{eqnarray}
	with
	\begin{eqnarray}
	{\beta}^{(n)}_{\eta}=\left\{
	\begin{aligned}
	1,\ \mathrm{if}\ \hat{\eta}^{(n)}=\eta\\
	0,\ \mathrm{if}\ \hat{\eta}^{(n)}\neq \eta
	\end{aligned}
	\right.
	\nonumber
	\end{eqnarray}
	
	\item \textit{Repeat}: Repeat the \textit{assignment} step and \textit{update} step until the assignments do not change. 
	\begin{eqnarray}
	\mathbf{m}^{(\eta)}_t=\mathbf{m}^{(\eta)}_{t+1}
	\label{eq:16}
	\end{eqnarray}
\end{itemize}
\subsubsection{$k$MC-KNN Algorithm}  
\begin{figure}[t]
	\centering
	\includegraphics{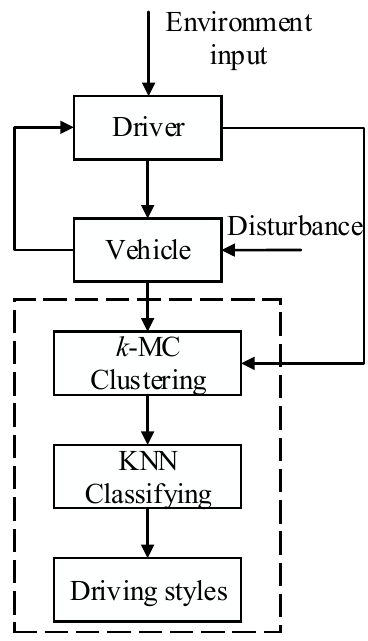}
	\caption{Schematic diagram of our developed $k$MC-KNN method.}
	\label{Fig.3}
\end{figure}

We developed a $k$MC-KNN algorithm, as shown in Fig. \ref{Fig.3}, to overcome the mentioned limitation of KNN in the classification procedure \cite{36}. With this purpose, we apply $k$-MC to search representatives of the whole training data to reduce the computational cost of KNN. There are three main steps of this recognition method, {including clustering, selecting and classifying. Clustering is to train the recognition model, selecting and classifying are to identify the pattern of new data.}
\begin{enumerate}
	\item \textit{Clustering}. The $k$-MC algorithm clusters the training samples of each category into $k$ sub-clusters. As a consequence, we obtain the recognition model.
	\item \textit{Selecting}. For a new input data, unlike the traditional KNN which recognizes its patterns with all the training data in each category, the $k$MC-KNN algorithm selects the subset with the largest similarity between the given data and the centres of each subsets in each category as the training samples by (\ref{eq:11})
	\begin{eqnarray}
	\hat{\eta}^{(*)}=\arg \min_\eta\Arrowvert \mathbf{x}^{(*)}-\hat{\mathbf{x}}^{(\eta)} \Arrowvert
	\label{eq:11}
	\end{eqnarray}
	where $\hat{\mathbf{x}}^{(\eta)}$ is the mean of the data in sub-clusters $\{(\mathbf{x}^{(\eta)},y_{j})\}^{k}_{\eta=1}$.
	
	\item \textit{Classifying}. Applying the KNN to classify the given data with the selected training samples. 
\end{enumerate}

The detailed procedure of $k$MC-KNN for driving style recognition is also shown in \textbf{Algorithm 1}. We shall learn the mapping between driving styles and driving data, formulated as $f:\mathcal{X}\rightarrow\mathcal{Y}$, where $\mathcal{X}=\{\mathbf{x}^{(n)}\}$ is a set of all collected driving data and $\mathcal{Y}=\{y^{(n)}\}_{n=1}^{N}$ is the set of driving style. 
\begin{algorithm}[t]
	\caption{:Algorithm for $k$MC-KNN} 
	{\bf Training}
	\begin{algorithmic}[1]
		\State  Given the labeled data set $\mathcal{M}=\{(\mathbf{x}^{(n)} ,y^{(n)} )\}^{N}_{n=1}$, $y^{(n)}\in\{y_j\}_{j=1}^{J}$, get  $  \{(\mathbf{\overline{x}}^{(n)},y^{(n)})\}^{N}_{n=1}$ using (\ref{eq:3}).
		\For{$j=1$ to $J$}
		\State Randomly choose $k$ instances $\{(\mathbf{m}^{(\eta)},y_{j})\}^{k}_{\eta=1}$ from $\{(\mathbf{\overline{x}}^{(n)},y_j)\}_{n}^{N_j}$.
		\While{$\mathbf{m}_t^{(\eta)}\neq\mathbf{m}_{t+1}^{(\eta)}$}
		\State Update $\mathbf{m}_t^{(\eta)} $ using (\ref{eq:10})
		\State Assign  $\mathbf{x}^{(n)}$ to  $ \hat{\eta}^{(n)}$ using (\ref{eq:9})
		\EndWhile
		\State Get the final $k$ sub-clusters $\{(\mathbf{x}^{(\eta)},y_{j})\}^{k}_{\eta=1}$
		\EndFor
		\State Get the $J\times k$ sub-clusters $\{(\mathbf{x}^{(\eta)},y_{j})\}^{k}_{\eta=1},j=1,2,\dots,J$
	\end{algorithmic}
	{\bf Testing}
	\begin{algorithmic}[1]
		\State Given the new data $\mathbf{x}^{(*)}$, we get  $\mathbf{\overline{x}}^{(*)}$ using  (\ref{eq:3}). 
		\For {$ j=1 $ to $J$}
		\State Select the nearest sub-cluster  $\{(\mathbf{x}^{(\eta^\ast)},y_{j})\}$ using (\ref{eq:11}) 
		\EndFor
		\State Get the selected $J$ sub-clusters $\{(\mathbf{x}^{(\eta^\ast)},y_{j})\}_{j=1}^{J}$
		\For{$ n=1 $ to $N$}
		\State Calculate  $\mathrm{SIM}(\mathbf{x}^{(n)},\mathbf{x}^{(*)})$ using (\ref{eq:6})
		\EndFor
		\State Choose the $K=\sqrt{N}$ samples with minimum similarity
		\State Judge $\mathbf{x}^{(*)}$ to $\hat{j}^{(*)} $ using (\ref{eq:13})
		\State \Return $\hat{j}^{(*)}$
	\end{algorithmic}
\end{algorithm}

\section{Experiments and Data Collection}
\subsection{Lane-Change Scenarios}

\begin{figure}[t]
	\centering
	
	\includegraphics[width=3.4in]{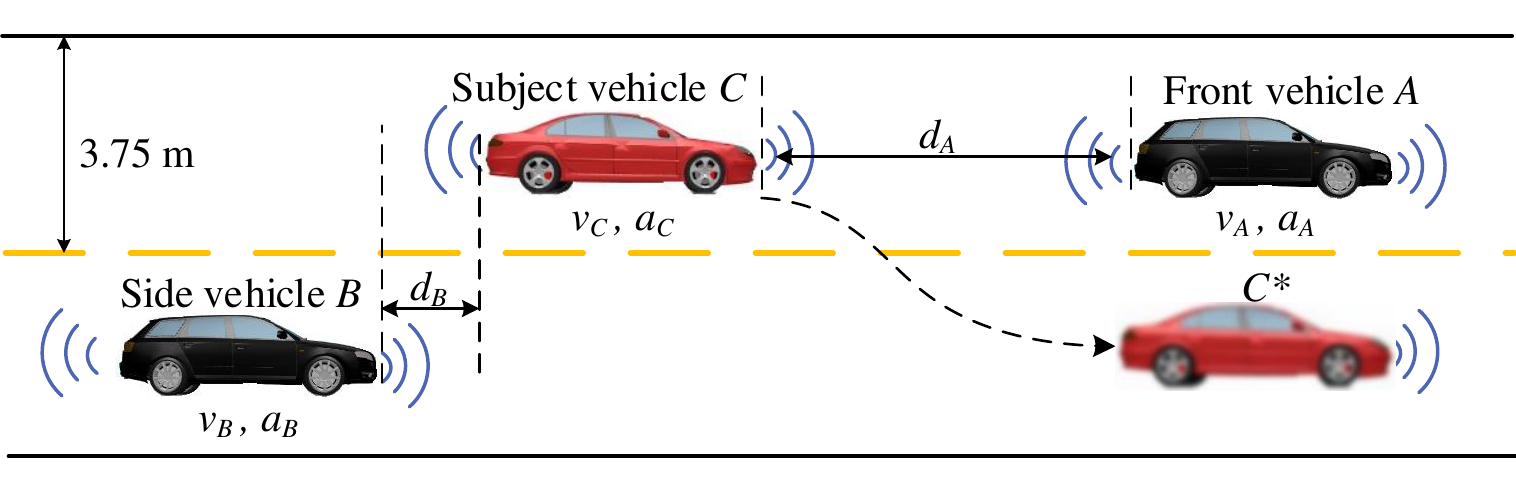}
	\hfil
	\caption{Specified lane changing scenarios.}
	\label{Fig.4}
\end{figure}


Among various driver behaviors, the lane-change maneuver occurs most frequently in real traffic \cite{39}. Drivers should be fully aware of the driving situation changes in order to make a safe decision and take a correct action when changing lanes. Completing a lane change task mainly requires three stages\cite{nilsson2016if}: \textit{determining} whether a lane-changing maneuver is desirable, \textit{selecting} the inter-vehicle traffic gap and initiation time, and \textit{planning} the longitudinal and lateral trajectory. The dynamic environment involved with surrounding vehicles is one of the main factors that influence the driver's decision-making, including chance determination and selection for lane changes. Accordingly, in order to show the efficiency of our proposed method to classify and recognize drivers' decision-making style, we conduct and analyze a typical lane-change scenario with three vehicles involved, as shown in Fig. \ref{Fig.4}.

In the driving scenario, the surrounding vehicles $ A $ and $ B $ drive straight at the speed of 40$\sim$60 km/h with a distance ($ d_A+d_B $) of 20$\sim$40 m, and then the subject vehicle $ C $ changes lanes between vehicles $ A $ and $ B $. A dynamic traffic environment was designed, allowing the vehicles $ A $ and $ B $ to accelerate and decelerate to maintain the distance of around 30 m. When the distance is greater than 30 m, the front vehicle $ A $ will brake slowly; meanwhile, the side vehicle $ B $ will increase the throttle opening. A complete lane change procedure was achieved when the driver steers the vehicle from the left lane to the center of the right lane, as shown from the vehicle $ C $ to the position of vehicle $ C^* $ in Fig. \ref{Fig.4}. All the involved vehicles were driving on a two-lane motorway with enough length to ensure that the driver can complete the lane change task. The lane width was set to 3.75 m, according to the Chinese national standard. All vehicles were equipped with vehicle-to-vehicle (V2V) capability in our simulation environment. The distance $d_A $ (or $ d_B$) between the vehicle $ A $ (or $ B $) and the subject vehicle $ C $ was recorded through V2V communication. The subject vehicle $ C $ also received the speeds (denoted as $v_A$ and $v_B$)  and accelerations (denoted as $a_A$  and $a_B$) of the surrounding vehicles $ A $ and $ B $.

\subsection{Feature Selection}

\begin{figure}[t]
	\centering
	\includegraphics[width=3.4in]{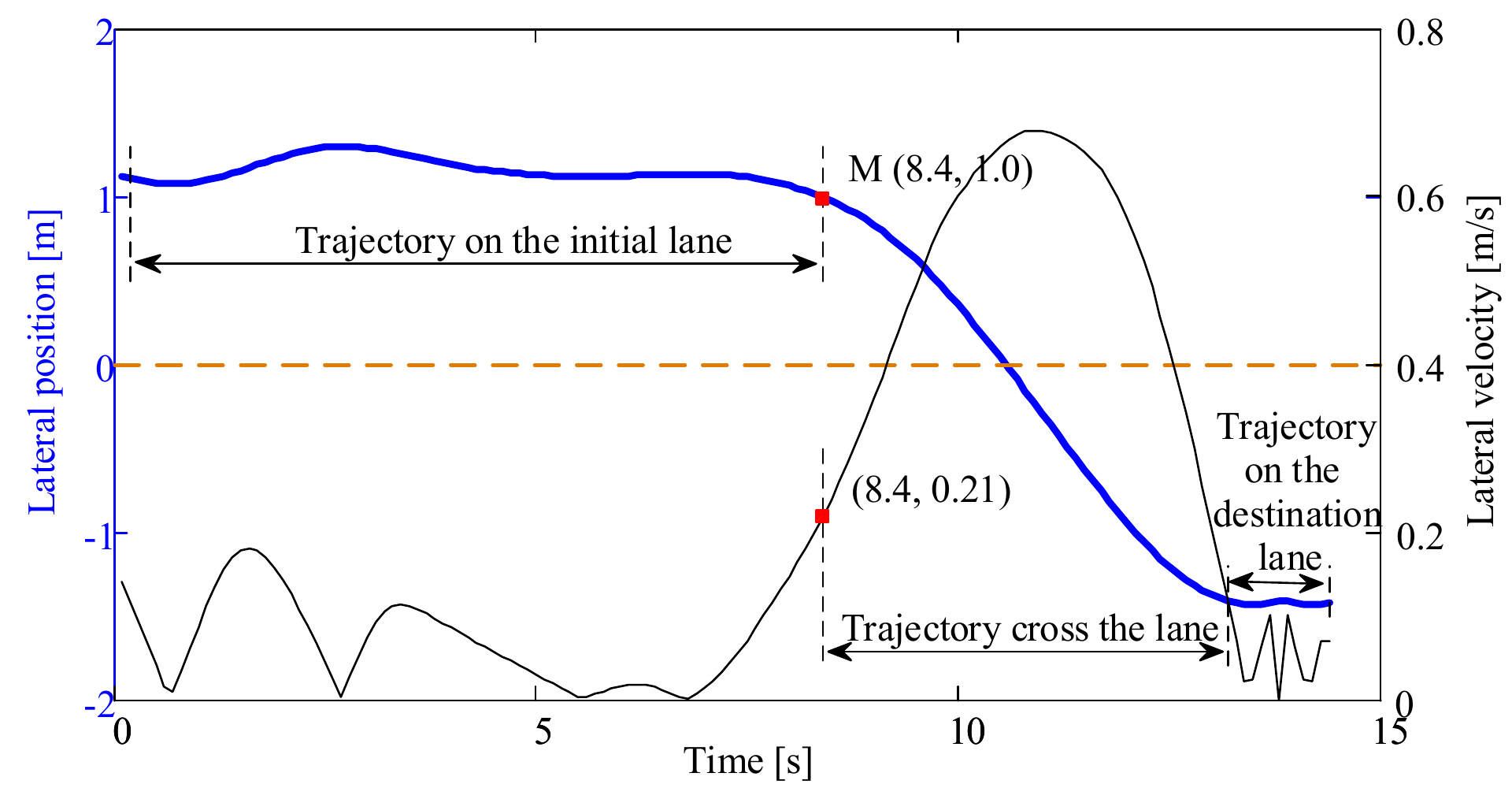}
	\caption{Example of lane-changing behavior.}
	\label{Fig.5}
\end{figure}

Feature selection is very important for driving style classification and recognition, which should allow pattern vectors to belong to different categories, so that they occupy compact and disjoint regions as much as possible in a specified feature space \cite{41}.  From a geometric point of view, the lane change behavior is a modification in the lateral position of the vehicle relative to  the current driving lanes, and can be divided into 3 segments (Fig. \ref{Fig.5}) \cite{42}: straight trajectory on the initial lane, trajectory across the line, and trajectory on the destination lane. In the first segment, the driver keeps observing the position, speed, and acceleration of the front vehicle $ A $ and the side vehicle $ B $, then decides whether or not to change lanes. 

Human drivers have different decision-making thresholds regarding when and whether to change lanes, which is essentially influenced by their surroundings, perceptible relative changes of environments, and their internal models\cite{Asaithambi2017Overtaking,Hou2012A}. Therefore, the relative change in information was selected to characterize driver's decision of changing lanes\cite{Hou2014Modeling}, including the distance between the front vehicle and the subject vehicle ($d_A$), the distance between the side vehicle and the subject vehicle ($d_B$), the speed difference between the front vehicle and the subject vehicle ($ v_{AC}=v_A-v_C$) and the speed difference between the side vehicle and the subject vehicle ($ v_{BC}=v_B-v_C$). Besides, drivers also prefer different levels of acceleration and deceleration when changing lanes \cite{Moridpour2007Modelling}. 

According to the above discussions, we select three relative information as the feature parameters $\mathbf{x}=(\Delta d,\Delta v, \Delta a)$ to characterize the driver's decision-making style during the lane-changing procedure, including the distance difference $\Delta d$ between $d_A$ and $d_B$, the relative speed difference $\Delta v$ between $v_{AC}$ and  $v_{BC}$ , and the relative acceleration difference $\Delta a$ between $a_{AC}=a_A-a_C$ and  $a_{BC}=a_B-a_C$, discussed as follows.

\begin{figure}[t]
	\centering
	\subfloat[$v_C=(v_B+v_A)/2$]{\includegraphics[width=3.4in]{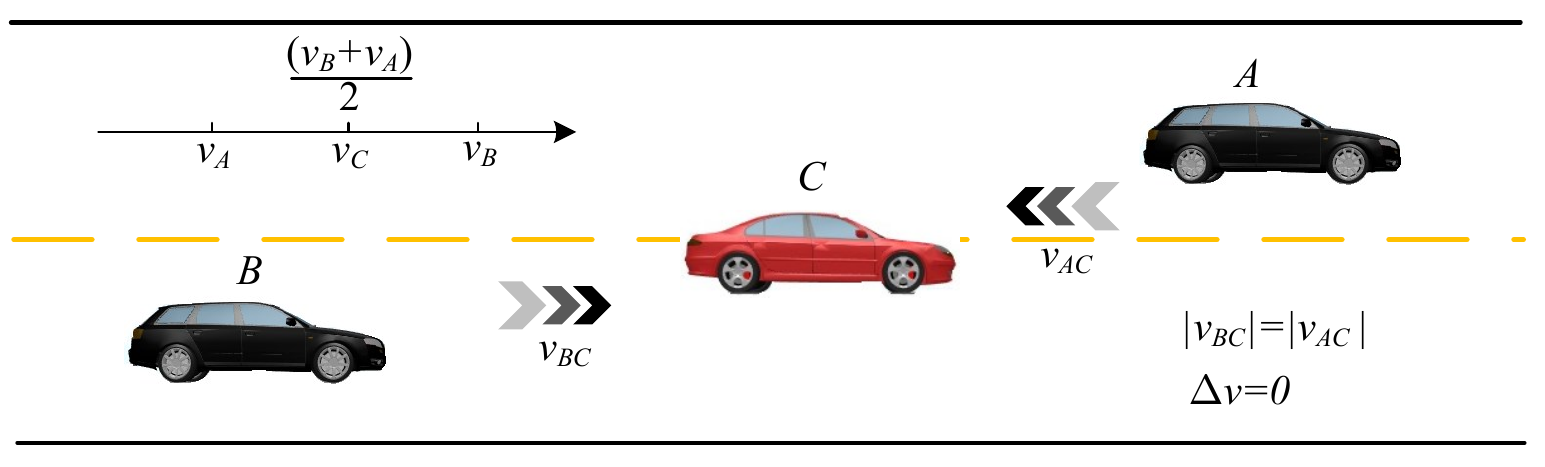}}
	\hfil
	\subfloat[$v_C<(v_B+v_A)/2$]{\includegraphics[width=3.4in]{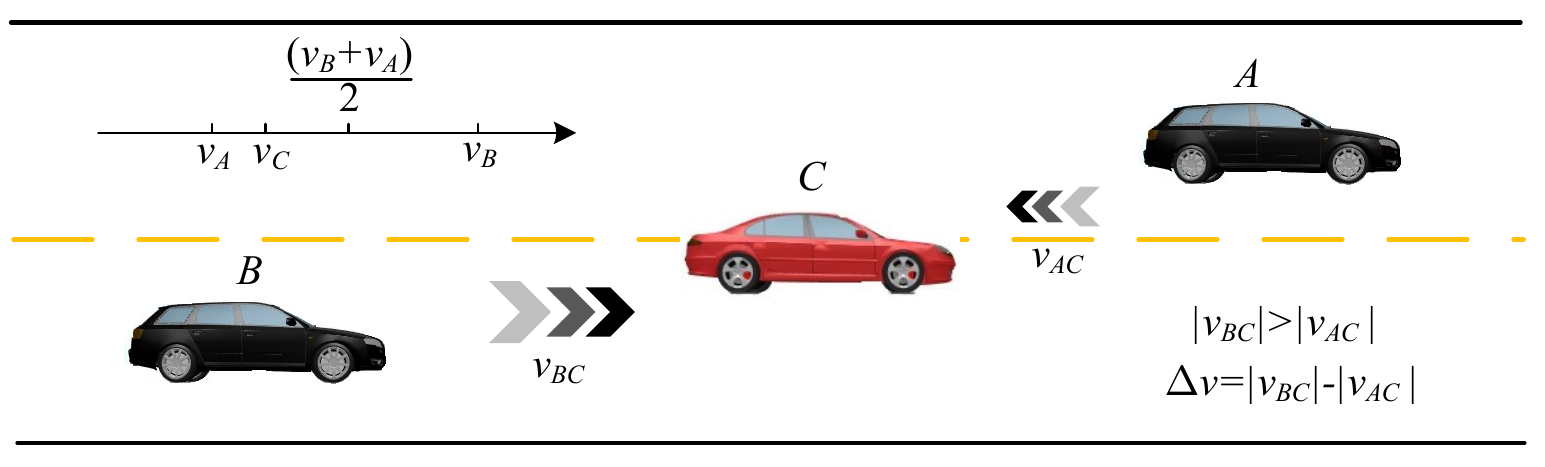}}
	\hfil
	\subfloat[$v_C>(v_B+v_A)/2$]{\includegraphics[width=3.4in]{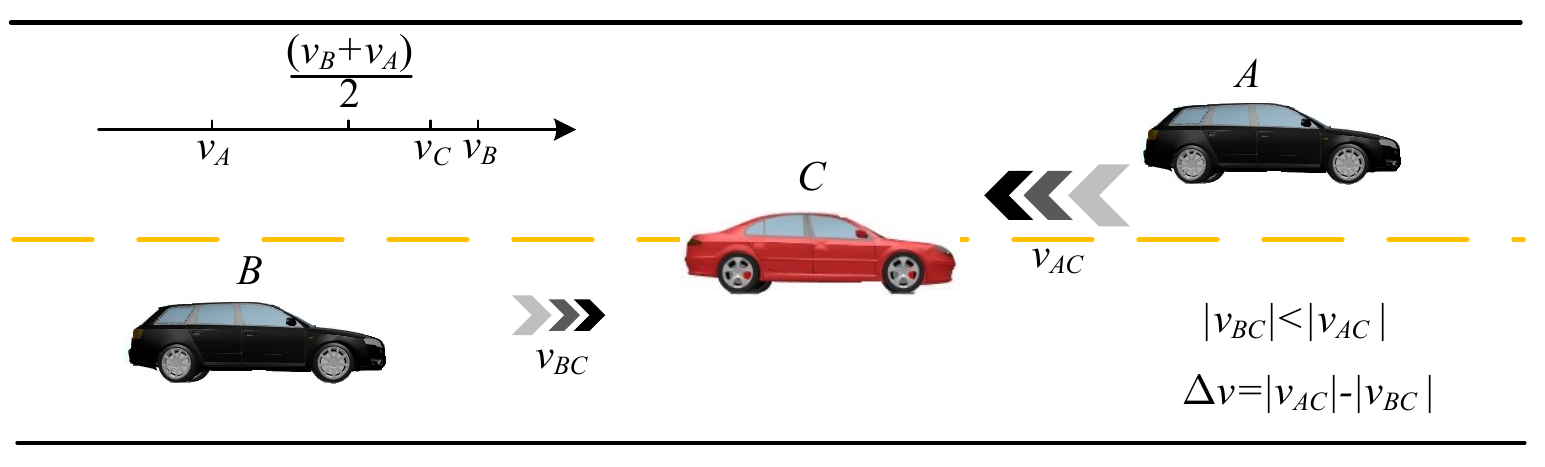}}
	\hfil
	\subfloat[$v_C<v_A$]{\includegraphics[width=3.4in]{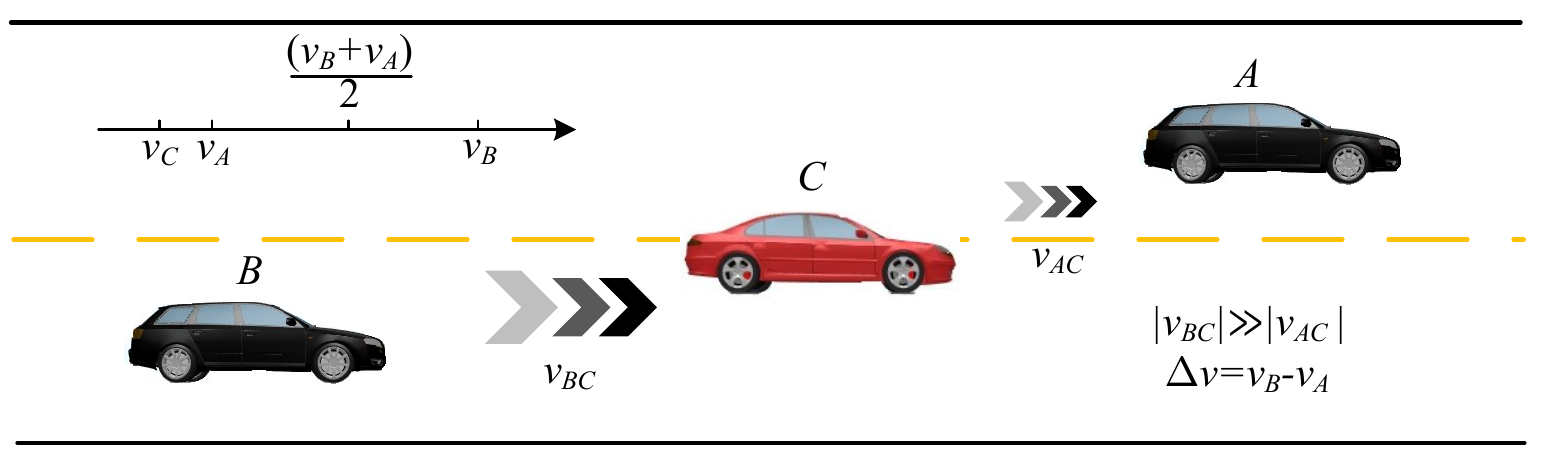}}
	\hfil
	\subfloat[$v_C>v_B$]{\includegraphics[width=3.4in]{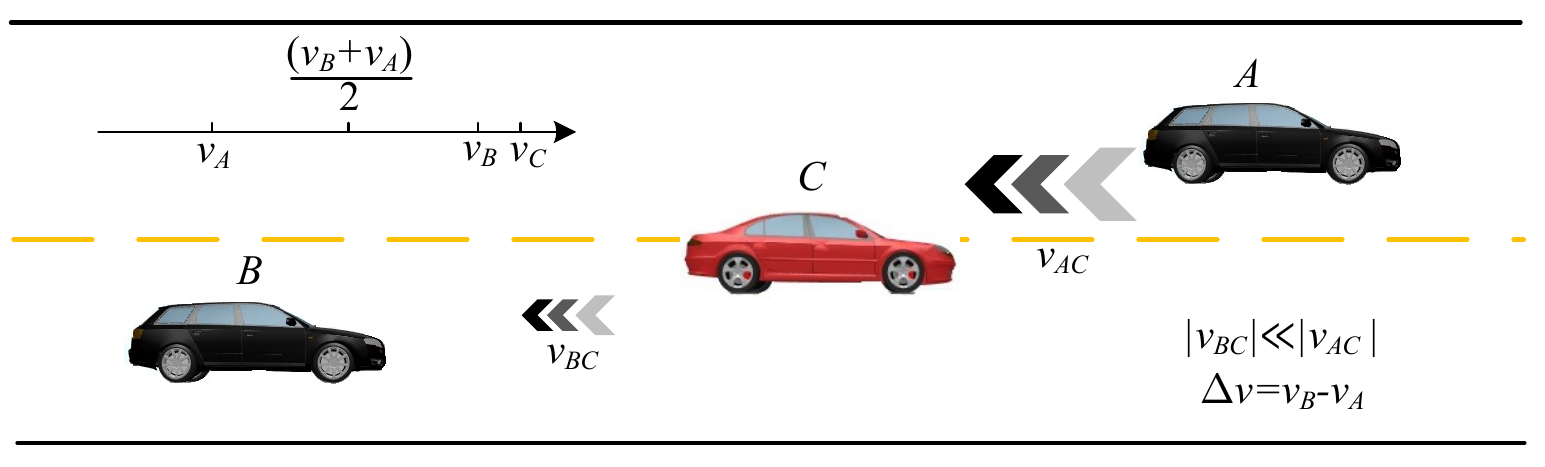}}
	\hfil
	\caption{Illustration of the relative speed difference when $v_A<v_B$.}
	\label{Fig.13}
\end{figure}

\begin{enumerate}
	\item \textit{Relative Distance Difference} ($\Delta d=|d_A-d_B|$): A greater distance difference indicates the subject vehicle is more close to the front vehicle or the side vehicle. Drivers who prefer a greater distance difference are more likely to make aggressive decisions for changing lanes.
	\item \textit{Relative Speed Difference} ({$ \Delta v=||v_{AC}|-|v_{BC}||$}): A higher relative speed difference heralds that the subject vehicle approaches the front vehicle or side vehicle with a higher speed. Drivers who prefer a large speed difference when changing lanes would be aggressive. 
	\item \textit{Relative Acceleration Difference} ($ \Delta a=||a_{AC}|-|a_{BC}||$): A larger acceleration difference indicates a more dangerous situation. Drivers who prefer a large acceleration would be treated as aggressive.
\end{enumerate}

In order to intuitively understand the relationship between the selected features and driving style, we visualize different typical cases of $ \Delta v $ when $v_A<v_B$, in Fig. \ref{Fig.13}. 
\begin{itemize}
	\item If $v_C=(v_B+v_A)/2$ (Fig. \ref{Fig.13}a), then $|v_{BC}|=|v_{AC}|$ and $\Delta v=0$, which indicates that the front and side vehicles are both approaching to the subject vehicle $C$ equally.
	\item If $v_C<(v_B+v_A)/2$ (Fig. \ref{Fig.13}b), then $|v_{BC}|>|v_{AC}|$ and $\Delta v=|v_{BC}|-|v_{AC}|$, which indicates that the side vehicle $B$ is approaching the subject vehicle $C$ faster than the front vehicle $A$. Therefore drivers in the case of $\Delta v=|v_{BC}|-|v_{AC}|$ drive more aggressively than in the case of $\Delta v=0$.  Analogously, the case of $\Delta v=|v_{AC}|-|v_{BC}|$ (Fig. \ref{Fig.13}c) is also more dangerous than the case of $\Delta v=0$, indicating the driver behave more aggressively.
	\item If $|v_{C}|<|v_{A}|$ (Fig. \ref{Fig.13}d), we have $|v_{BC}|\gg|v_{AC}| $ and $\Delta v=v_B-v_A$, which indicates that the side vehicle $B$ is approaching the subject vehicle $C$ faster than itself in the case of $\Delta v=|v_{BC}|-|v_{AC}|$. Therefore drivers in the case of $|v_{C}|<|v_{A}|$ drive more aggressively than in the case of $v_C<(v_B+v_A)/2$. Analogously, the case of  $|v_{C}|>|v_{B}|$ (Fig. \ref{Fig.13}e) is also more dangerous than the case of $v_C>(v_B+v_A)/2$, indicating the driver is more aggressive.

\end{itemize}
The principles of the relative distance difference and relative acceleration difference can also be interpreted in the same way as what we do for the relative speed difference.


\begin{figure}[t]
	\centering
	\subfloat[]{\includegraphics[width=3.4in]{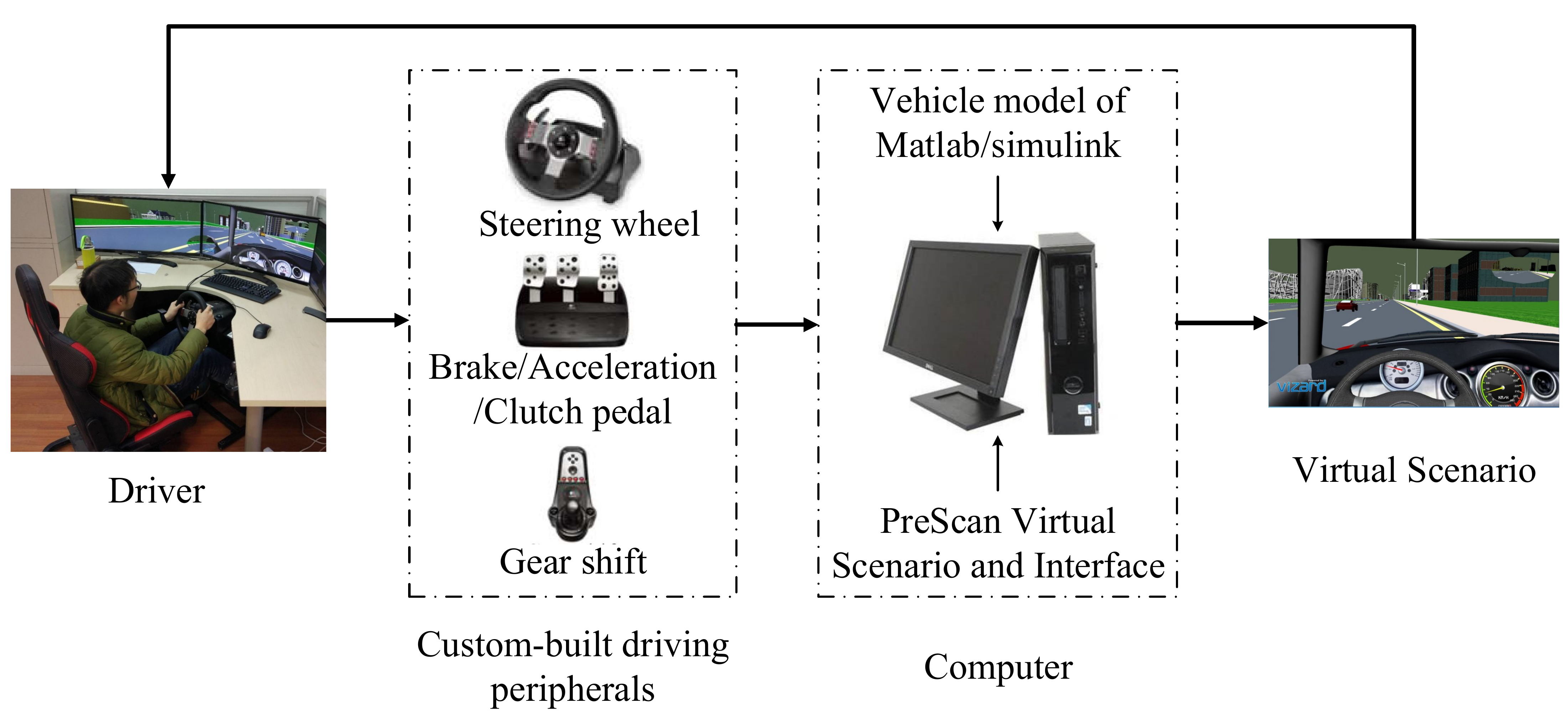}}
	\hfil
	\subfloat[]{\includegraphics[width=0.223\textwidth]{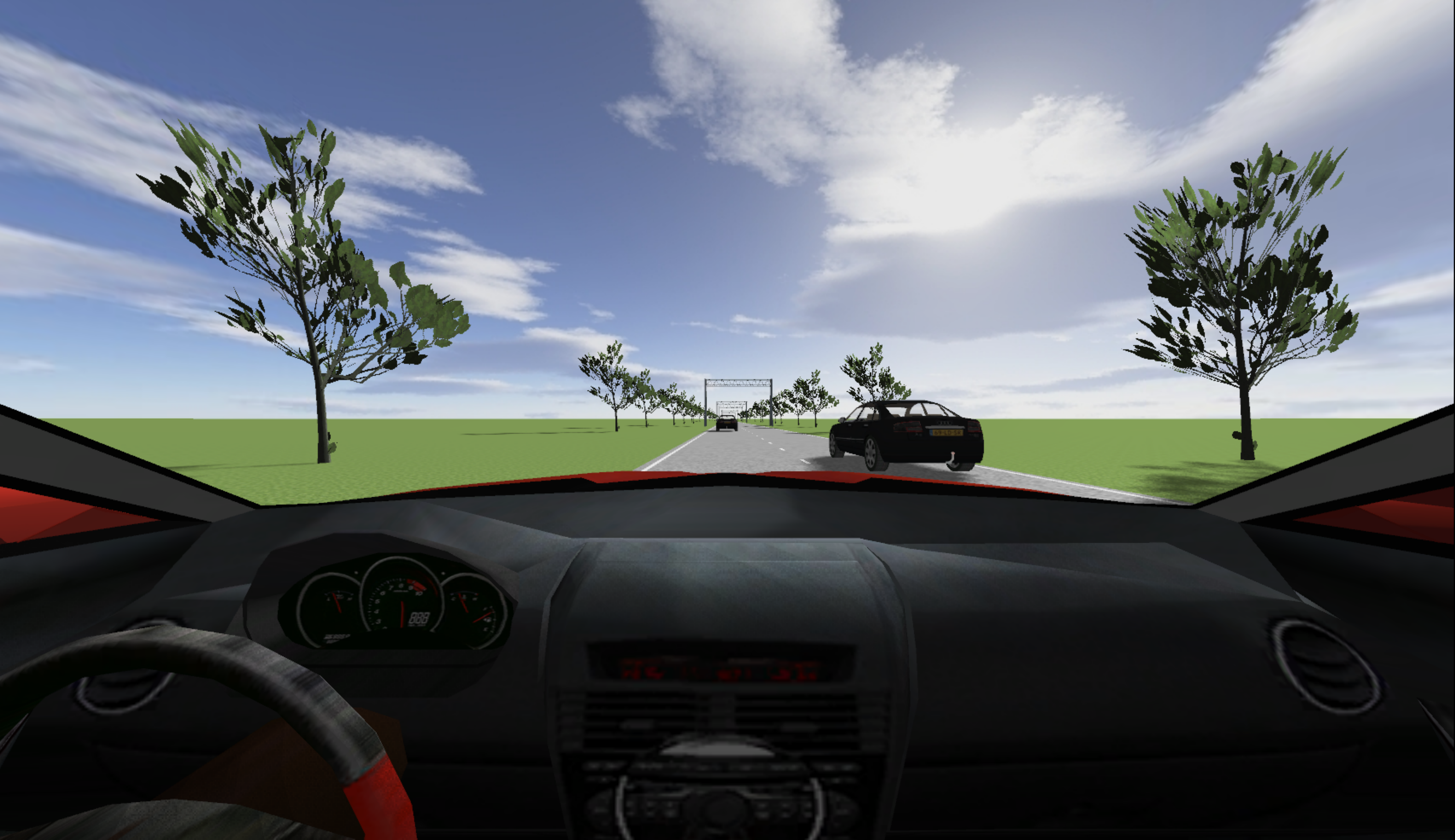}}
	\hfil
	\subfloat[]{\includegraphics[width=0.247\textwidth]{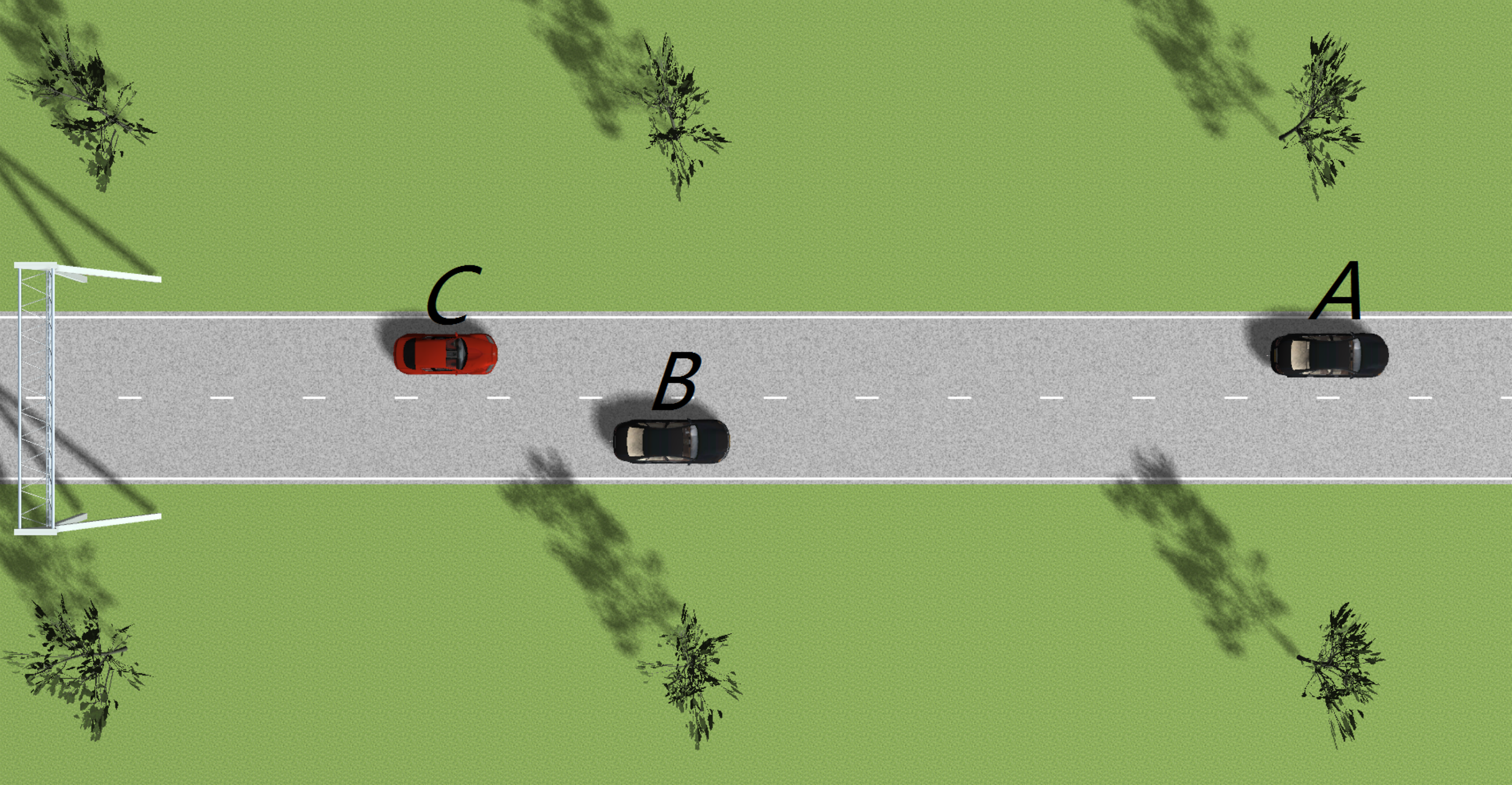}}
	\hfil
	\caption{(a) Schematic diagram of driving simulator. (b) The driver view of scenario \textit{A} in the PreScan. (c) The top view of scenario \textit{A} in the PreScan}
	\label{Fig.6}
\end{figure}

\subsection{Driving Simulator and Data Collection}
The training and testing data were collected in a driving simulator (Fig. \ref{Fig.6}). The driving simulator consists of four main parts: human driver, operation input equipment, vehicle dynamics model, and the virtual environment display. The custom-built driving peripherals, including steering wheel, brake/acceleration/clutch pedal, and gear shift handle, were utilized to collect the driver's operating signals such as steering wheel angle, brake pedal position, and throttle opening. A bicycle-vehicle dynamics model was built using Matlab/Simulink. The virtual scenarios, including the vehicle, roads, and traffic facilities were designed through PreScan software.

\begin{figure}[t]
	\centering
	\subfloat[]{\includegraphics[width=3.4in]{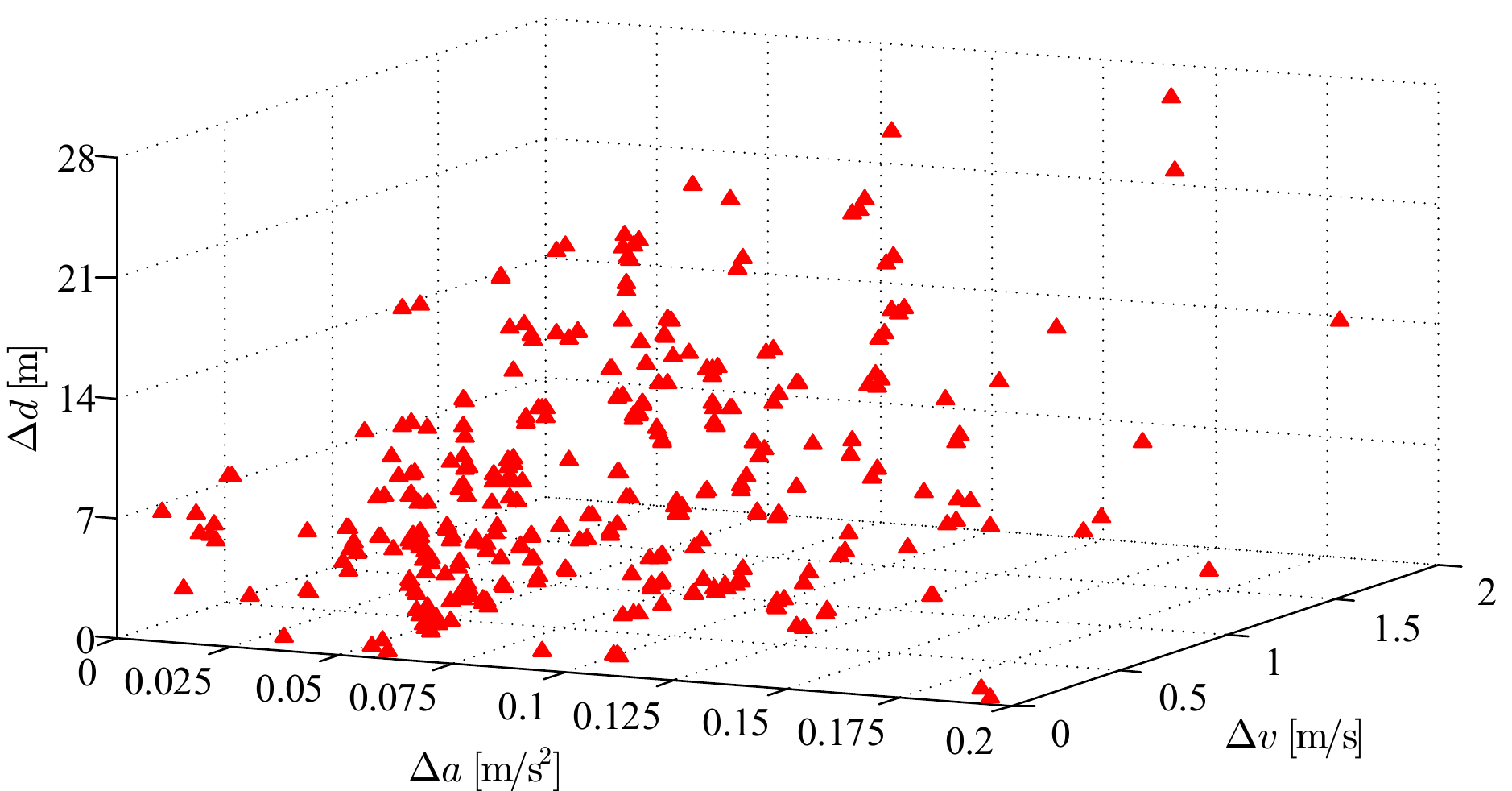}}
	\hfil
	\subfloat[]{\includegraphics[width=0.15\textwidth]{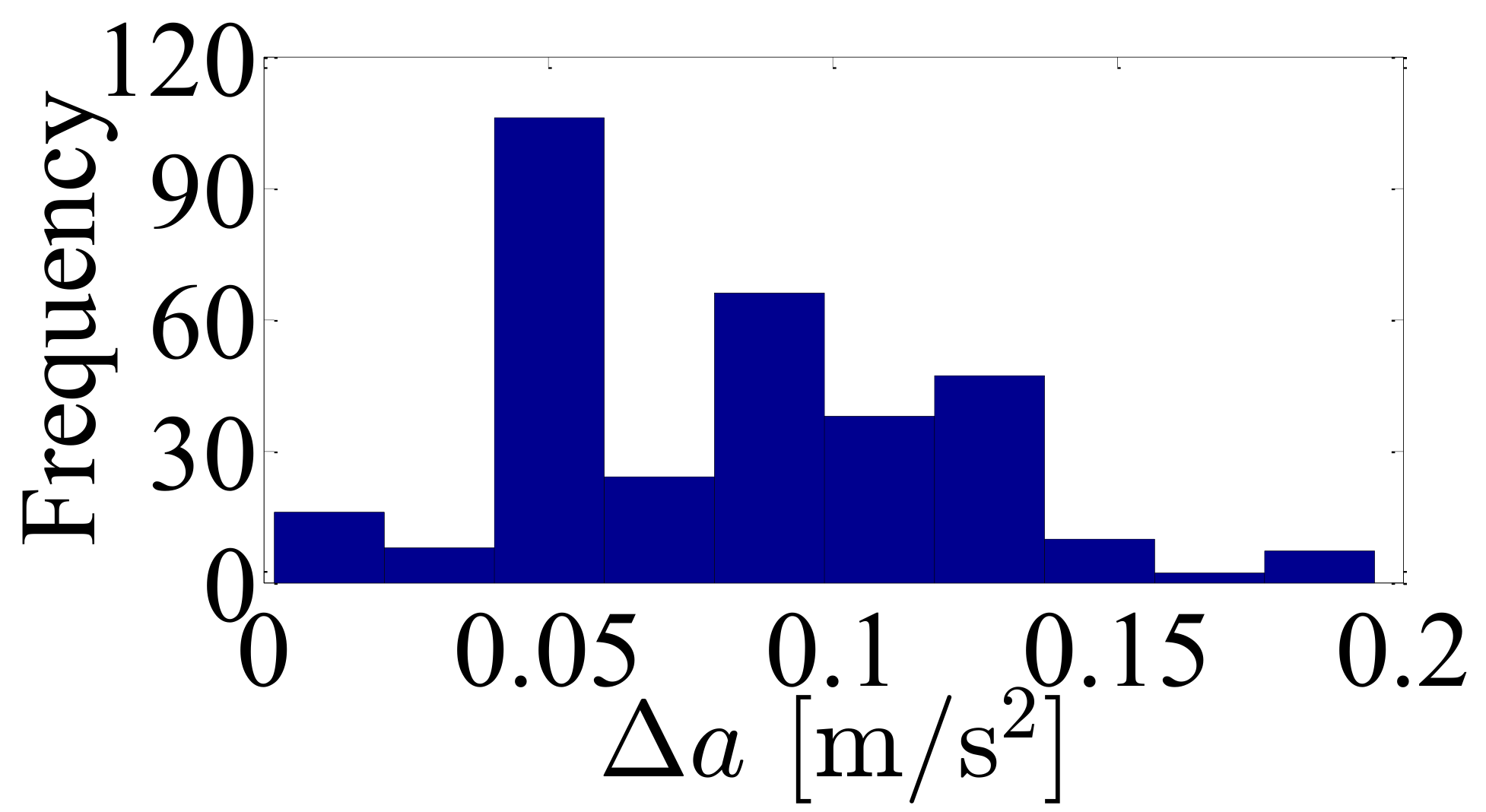}}
	\hfil
	\subfloat[]{\includegraphics[width=0.15\textwidth]{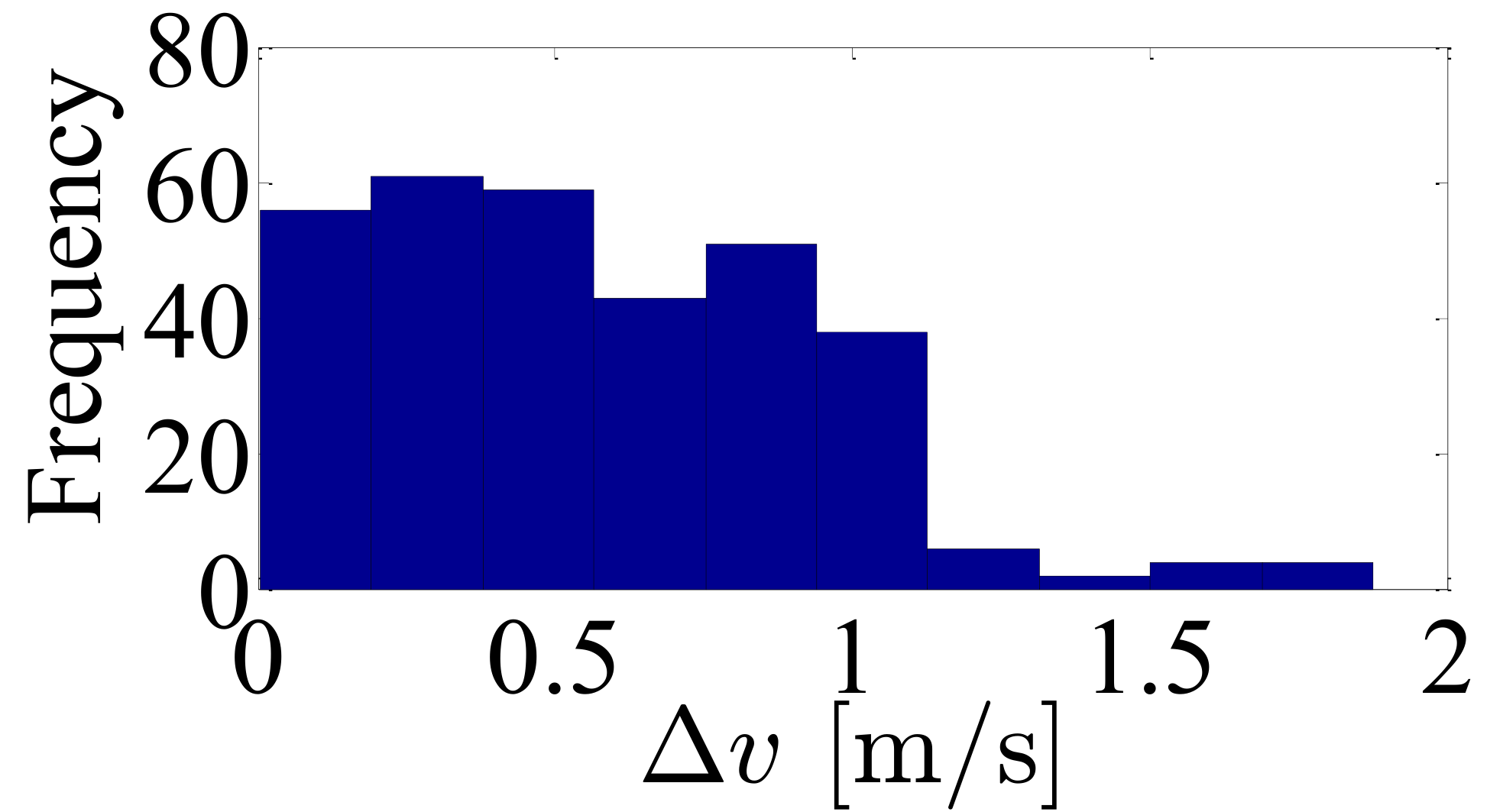}}
	\hfil
	\subfloat[]{\includegraphics[width=0.15\textwidth]{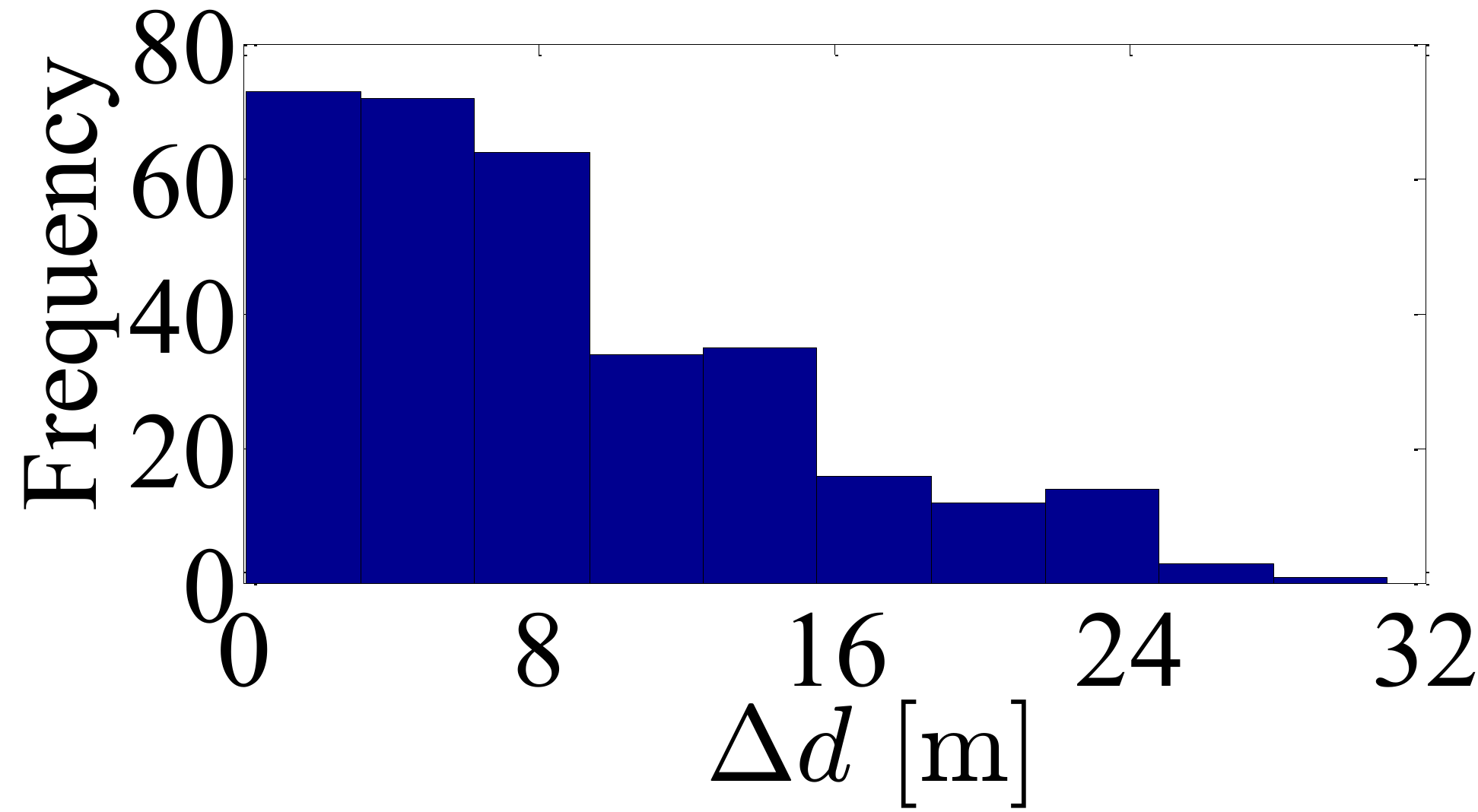}}
	\hfil
	\caption{The collected driving data from all drivers and their distribution.}
	\label{Fig.7}
\end{figure}

Totally, 16 subjects (12 males and 4 females) participated our experiment as volunteers, with a minimal of 22 years old and a maximal of 28 years old. All of the participants had held driver licenses for a minimal of 2 years. Each driver executed 25 trials of changing lanes repeatedly. Each driver was familiarized with the test course and the driving simulator before taking trials. During the trials, all drivers followed the rules: the secondary tasks such as talking with others, making or answering a telephone were forbidden; each participant rested 2 minutes before the next trial; all participants were in mentally and physically normal states; all participants manipulated the subject vehicle in their own driving style without any guidance.

All the collected data were time series, therefore we should define the specific decision-making moments of lane change in order to obtain high-quality training and testing data. It is defined as the moment that when the lateral velocity of the host vehicle $ C $ is up to $0.21 $ m/s which signifies discretionary lane-changing execution start \cite{47}, as illustrated by the point $ M $ in Fig. \ref{Fig.5}. Thus, the driving data at that moment was extracted as the feature data to characterize the driver's decision-making style in lane change scenarios. Fig. \ref{Fig.7}(a) shows the extracted experimental data of point $ M $. Figs. \ref{Fig.7}(b)-(d) show the distributions of different features. We can see that 1) the relative acceleration difference is not strictly subject to a uniform distribution and most data points gather around $0.05$ m/s$^2$ or $0.1$ m/s$^2$; 2) the data samples of relative velocity difference approximately fall in $[0, 1.2]$ m/s, and only a few of data samples are greater than $1.2$ m/s; and 3) the data of relative distance difference range in $[0, 32]$ m.

\section{Experiment Result Analysis and Evaluation}
This section will analyze and evaluate the experiment results of our developed $k$MC-KNN method by comparing with the traditional methods, including KNN and SVM.

\subsection{Clustering Result Analysis and Evaluation}
\subsubsection{Analysis}
\begin{figure}[t]
	\centering
	\subfloat[]{\includegraphics[width=0.24\textwidth]{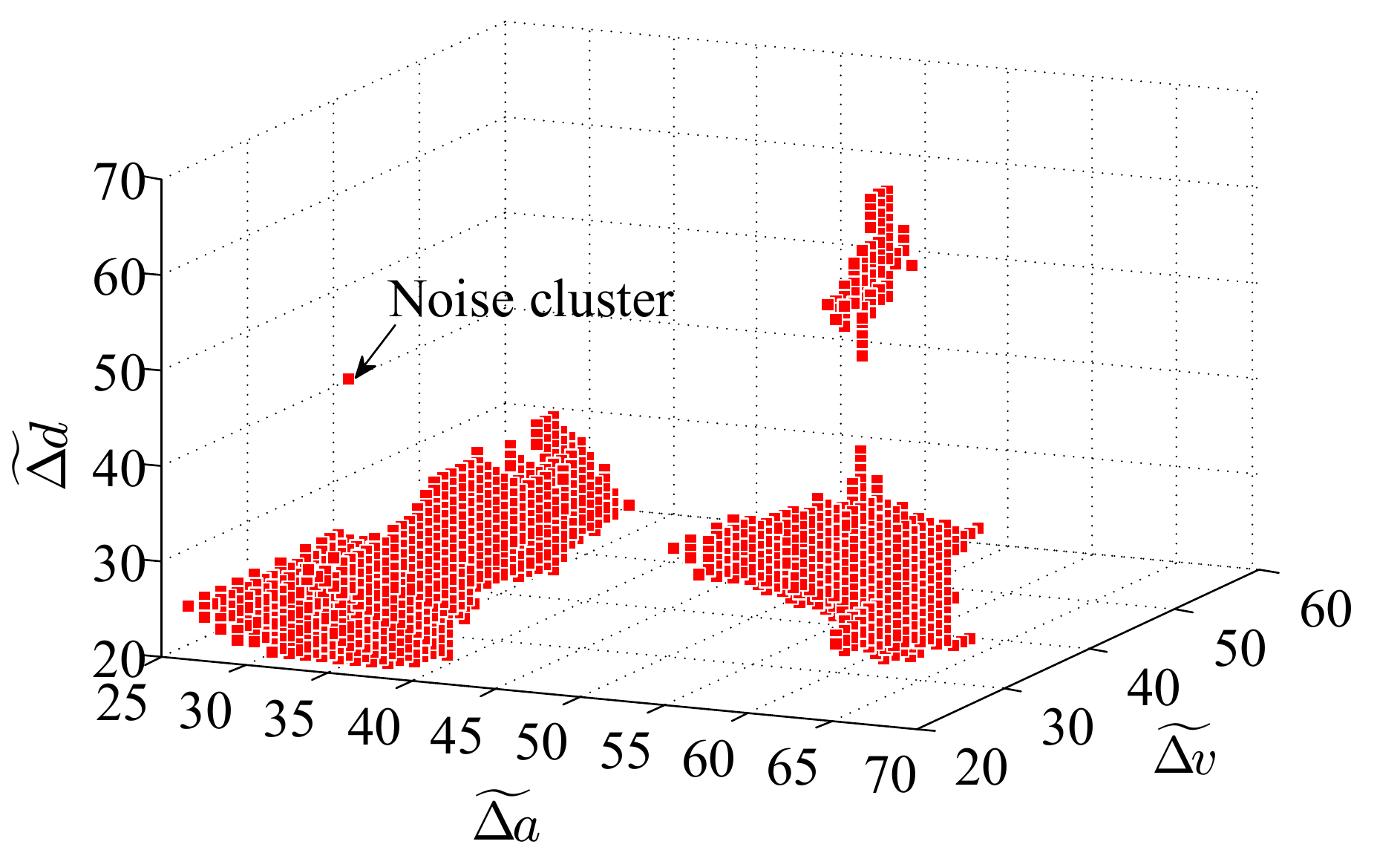}}
	\hfil
	\subfloat[]{\includegraphics[width=0.24\textwidth]{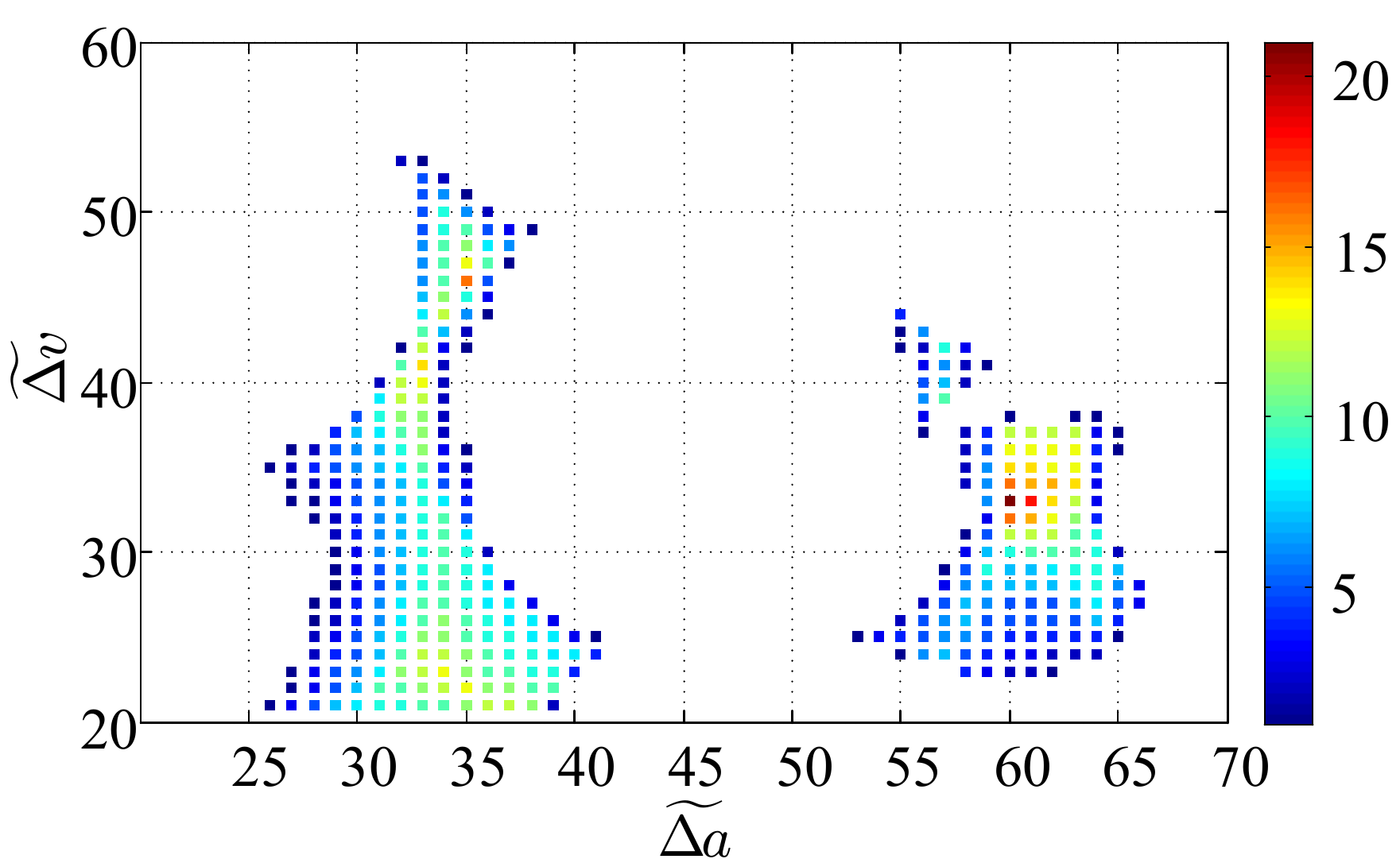}}
	\hfil
	\subfloat[]{\includegraphics[width=0.24\textwidth]{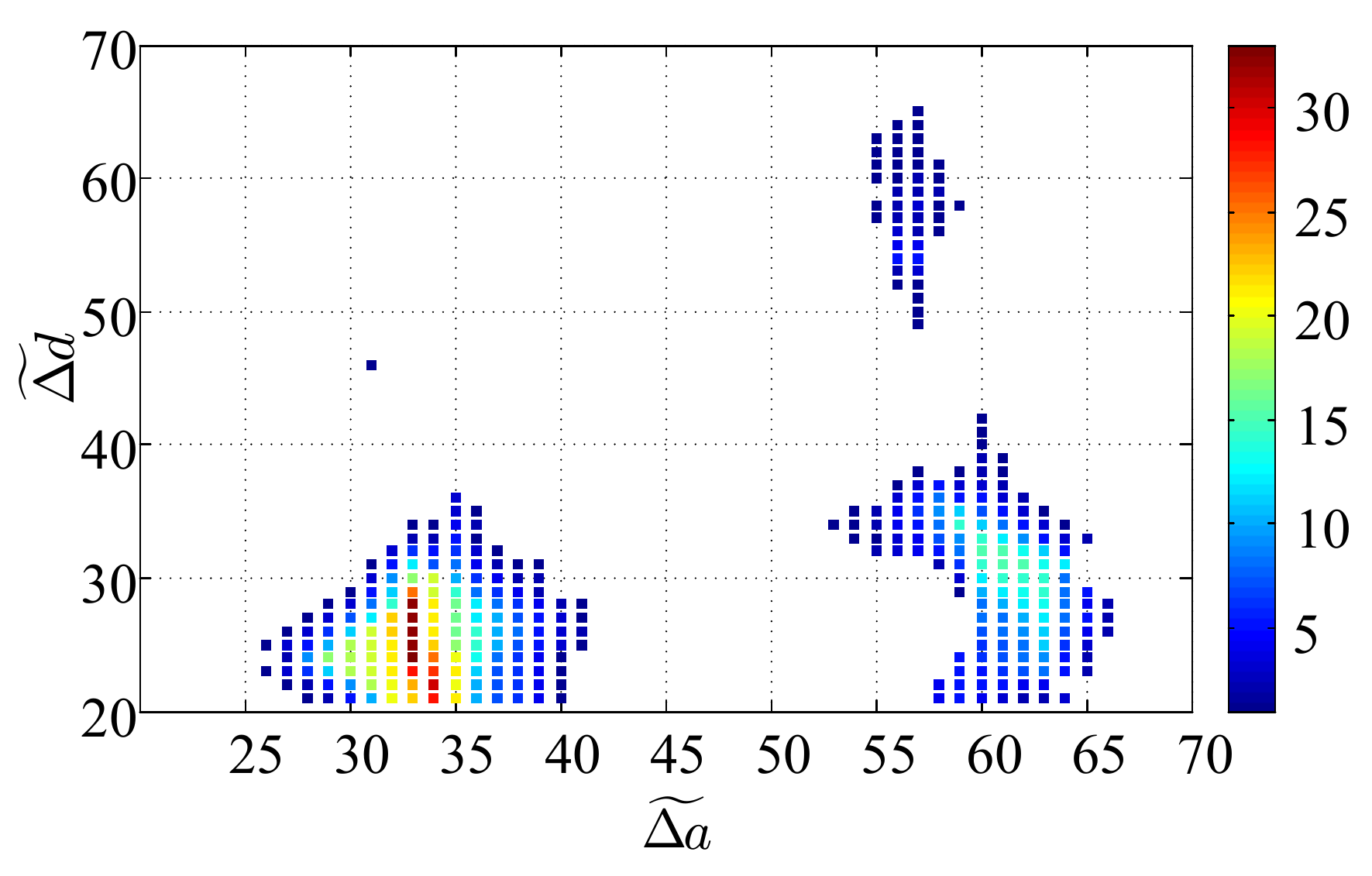}}
	\hfil
	\subfloat[]{\includegraphics[width=0.24\textwidth]{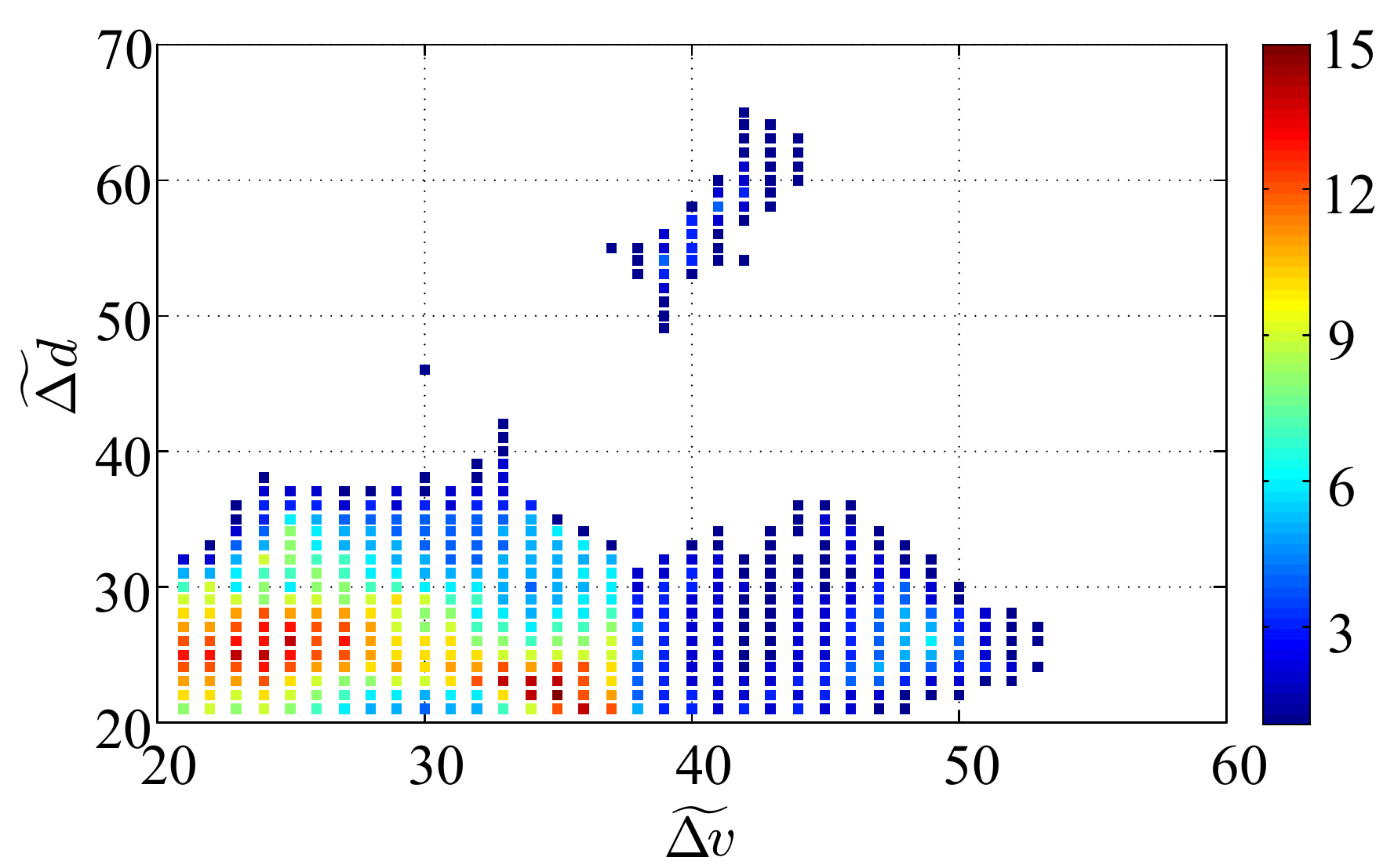}}
	\hfil
	\caption{Clustering results of using our proposed mathematical morphology-based method with dilation and erosion procedures.}
	\label{Fig.8}
\end{figure}

\begin{figure}[t]
	\centering
	\includegraphics[width=0.49\textwidth]{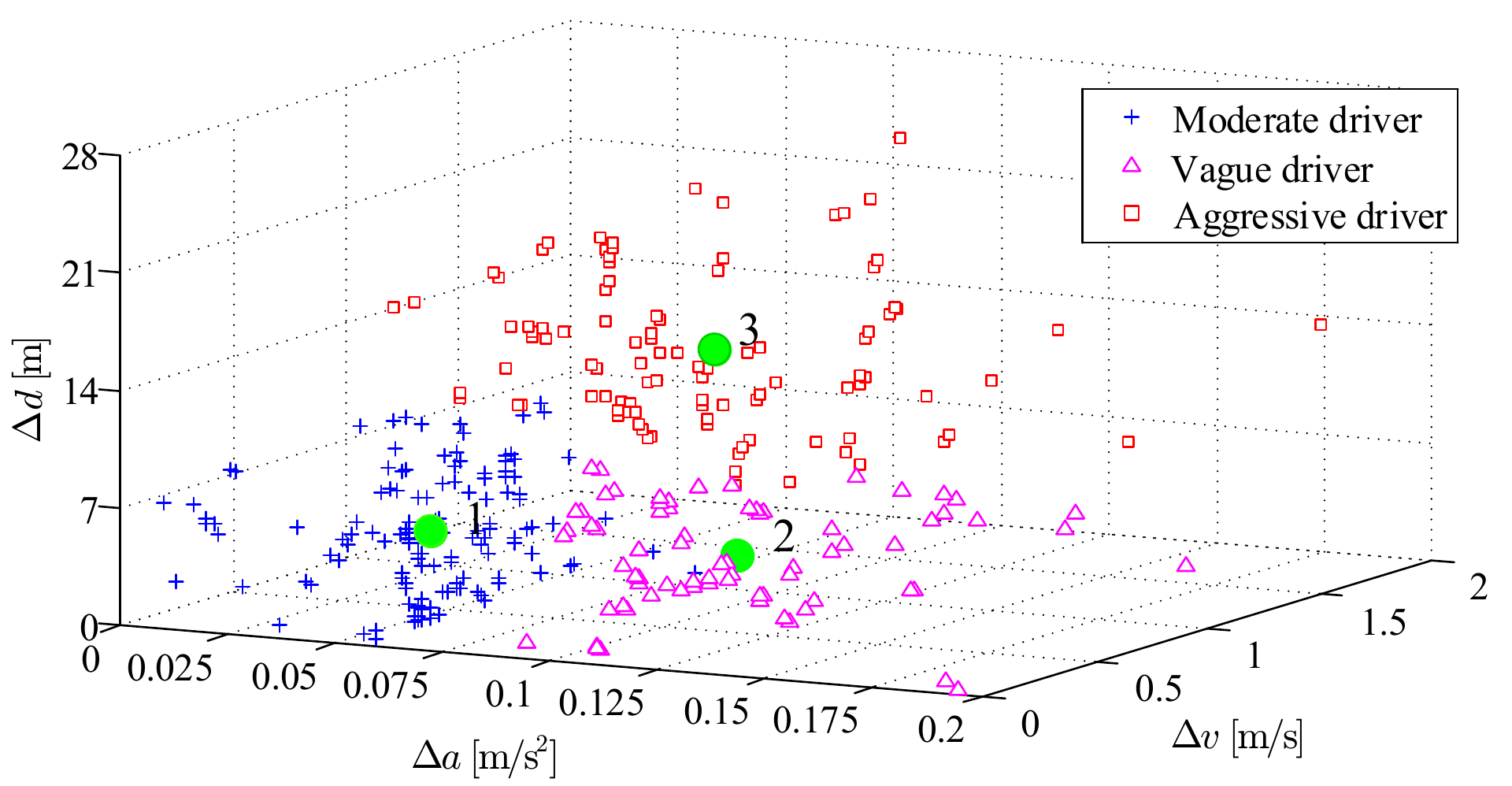}
	\caption{Clustering results for original data.}
	\label{Fig.9}
\end{figure}

The collected data are finally clustered into three decision-making groups and one noise group (Fig. \ref{Fig.8}) using the mathematical morphology-based clustering method with $q_{\Delta d}=q_{\Delta v}=q_{\Delta a}=100$ and $r=10$. The centers ($C_{\mathrm{mod}},C_{\mathrm{vag}},C_{\mathrm{agg}}$) and ranges of each cluster for each decision-making style are shown in Table \ref{Tab.1}. The raw driving data are assigned to these driving styles according to (\ref{eq:5}), as shown in Fig. \ref{Fig.9}. All training data are automatically labeled using the mathematical morphology-based approach with little effort of labeling data and little subjective interference. 


\begin{table*}[t]
	\renewcommand{\arraystretch}{1.3}
	\caption{Clustering Centers and Ranges of Each Driving Style Using Our Mathematical Morphology-Based Clustering Method}
	\centering
	\label{Tab.1}
	\begin{tabular}{c|c|c}
		\hline\hline
		Driving style & Cluster centers & Ranges\\
		\hline
		Moderate driver & $C_{\mathrm{mod}}=$ (0.0470, 0.4779, 4.6597) & (0.0020$ \sim $0.0831, 0.0036$ \sim $1.1273, 0.0090$ \sim $12.4460)\\
		Vague driver & $C_{\mathrm{vag}}=$ (0.1153, 0.5335, 4.4702) & (0.0850$ \sim $0.1951, 0.0360$ \sim $1.2958, 0.0090$ \sim $11.5193)\\
		Aggressive diver & $C_{\mathrm{agg}}=$ (0.1012, 0.6962, 15.727) & (0.0444$ \sim $0.1854, 0.0223$ \sim $1.8764, 6.2681$ \sim $30.9796)\\
		\hline\hline
	\end{tabular}
\end{table*}

The points with different shapes in Fig. \ref{Fig.9} represent different driving styles. Blue crosses represent drivers who prefer a low relative speed difference ($\leq1$ m/s), a small relative acceleration difference ($\leq 0.075$ m/s$ ^2 $), and a narrow relative distance difference ($\leq 10$ m) when making a lane-change decision. We tag these kind of drivers as \textit{moderate style} in decision-making. When the relative acceleration difference reaches a certain threshold ($0.08$ m/s$ ^2 $), the moderate driver has less preference to change lanes than other two types of drivers, which indicates the moderate driver is inclined to make a more conservative lane change. In addition, the moderate driver rarely drives the vehicle with a relative distance difference of larger than $10$ m. 

Red squares represent drivers who prefer a large relative distance difference ($\geq 10$m) in most cases, covering only a few points with small relative distance difference. This kind of drivers is categorized as \textit{aggressive style}. When $ \Delta d \geq 10$ m, the aggressive driver prefers a large relative acceleration difference ($\geq 0.05$ m/s$ ^2 $), which indicates that the aggressive driver prefers risky lane-changing maneuvers. Besides, the relative speed difference of the aggressive drivers is in a large range of $[0,2]$ m/s. 

Purple triangles represent drivers who prefer to change lanes with a relative acceleration difference in the range of $[0.08,0.2]$ m/s$ ^2 $. We categorize these kind of drivers under the \textit{vague style}. When the relative acceleration difference $ \Delta a\in[0.08,0.2]$ m/s$ ^2 $, the vague driver prefers a small relative speed difference ($\leq 1$ m/s) and a small relative distance difference ($\leq 10$ m). When $\Delta d\leq10$ m,  the vague drivers prefer a larger relative acceleration difference than the moderate driver. When $\Delta a\in[0.08,0.2]$ m/s$ ^2 $, the vague driver prefers a smaller relative distance difference than the aggressive driver.

Comparing the centers of three driving styles in Table \ref{Tab.1}, it can be concluded that for each variable, the moderate driver obtains a smaller value than the aggressive driver's counterpart. For vague drivers, only the relative distance difference is smaller than the aggressive drivers, and only the relative acceleration difference is larger than the moderate drivers.
\subsubsection{Evaluation}
To demonstrate the correctness of our proposed method, we compare it with the agglomerative hierarchical clustering (AHC) which is an important and well-established technique in unsupervised machine learning. AHC starts from the partition of the data set into singleton nodes and merges the current pair of mutually closest nodes into a new node step-by-step until there is one final node left, which comprises the entire data set. 

\begin{figure}[t]
	\centering 
	\includegraphics[width=3.4in]{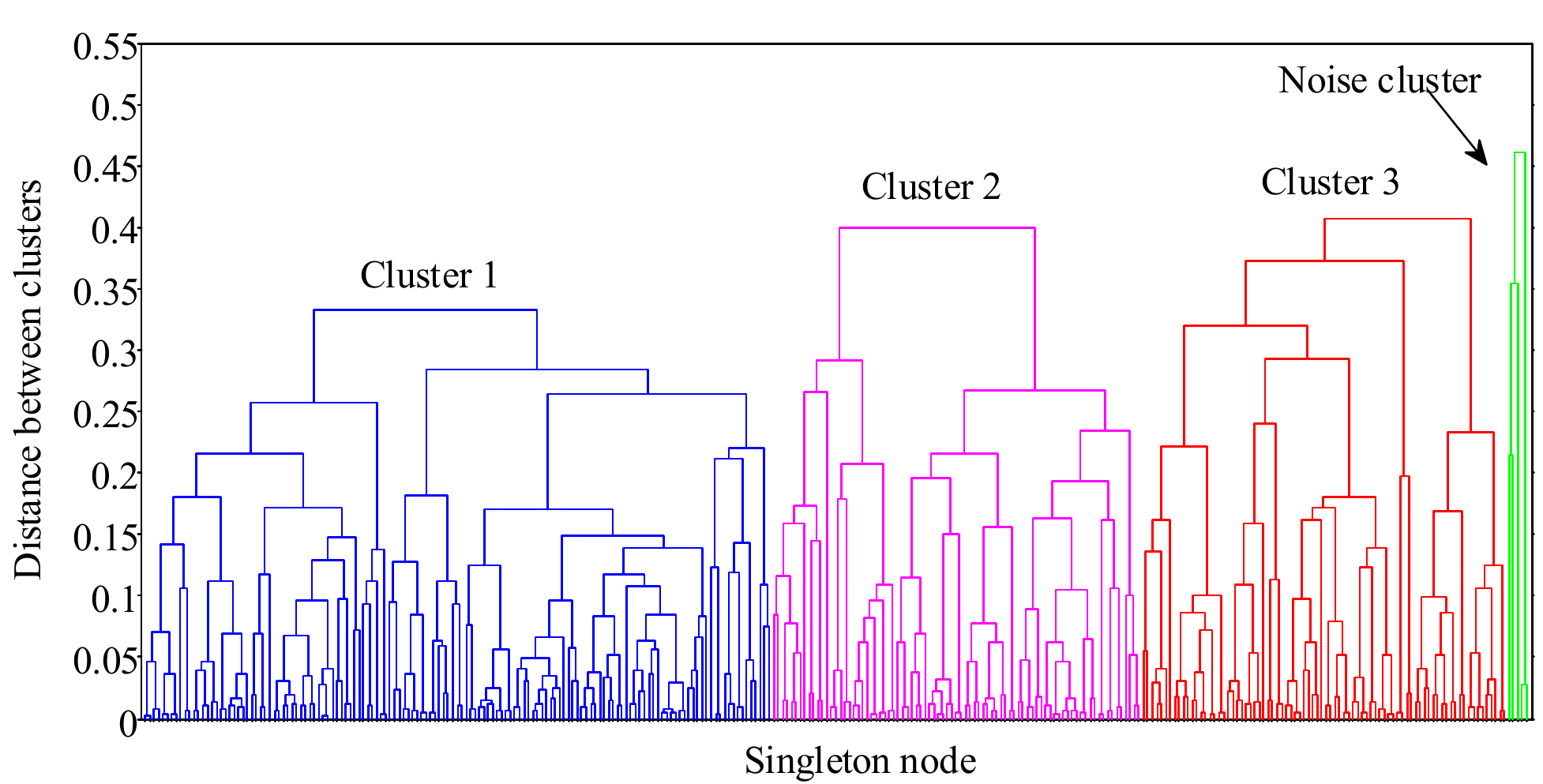}
	\caption{Clustering results of using the AHC method.}
	\label{Fig.10}
\end{figure}

Fig. \ref{Fig.10} presents the results of the final four clusters using AHC. We can see that the training data is classified into three main clusters and one noise cluster. The center and range of each cluster are shown in Table \ref{Tab.2}. Comparing the clustered centers in Table \ref{Tab.1} with those in Table \ref{Tab.2}, we can conclude that the cluster 1, cluster 2, and cluster 3 in Fig. \ref{Fig.10} are associated with the moderate, vague and aggressive decision-making styles in Fig. \ref{Fig.9}, respectively.

\begin{table*}[t]
	\renewcommand{\arraystretch}{1.3}
	\caption{Centers and Ranges of Each Cluster Using The AHC Method}
	\centering
	\label{Tab.2}	
	\begin{tabular}{c|c|c}
		\hline\hline
		Cluster \# & Centers & Ranges\\
		\hline
		Cluster 1 & $C_{\mathrm{mod}}^{'}=$ (0.0478, 0.4837, 4.9682) & (0.0020$ \sim $0.0767, 0.0123$ \sim $1.1284, 0.0903$ \sim $13.0704)\\
		Cluster 2 & $C_{\mathrm{vag}}^{'}=$ (0.1116, 0.6091, 5.0022) & (0.0847$ \sim $0.1648, 0.0360$ \sim $1.3468, 0.2208$ \sim $11.7478)\\
		Cluster 3 & $C_{\mathrm{agg}}^{'}=$ (0.1002, 0.5437, 16.6271) & (0.0446$ \sim $0.1516, 0.0266$ \sim $1.0668, 10.8425$ \sim $25.0622) \\
		\hline\hline
	\end{tabular}

\end{table*}
\subsection{Recognition Performance Analysis and Evaluation}
\begin{figure}[t]
	\centering
	\includegraphics[width=3.4in]{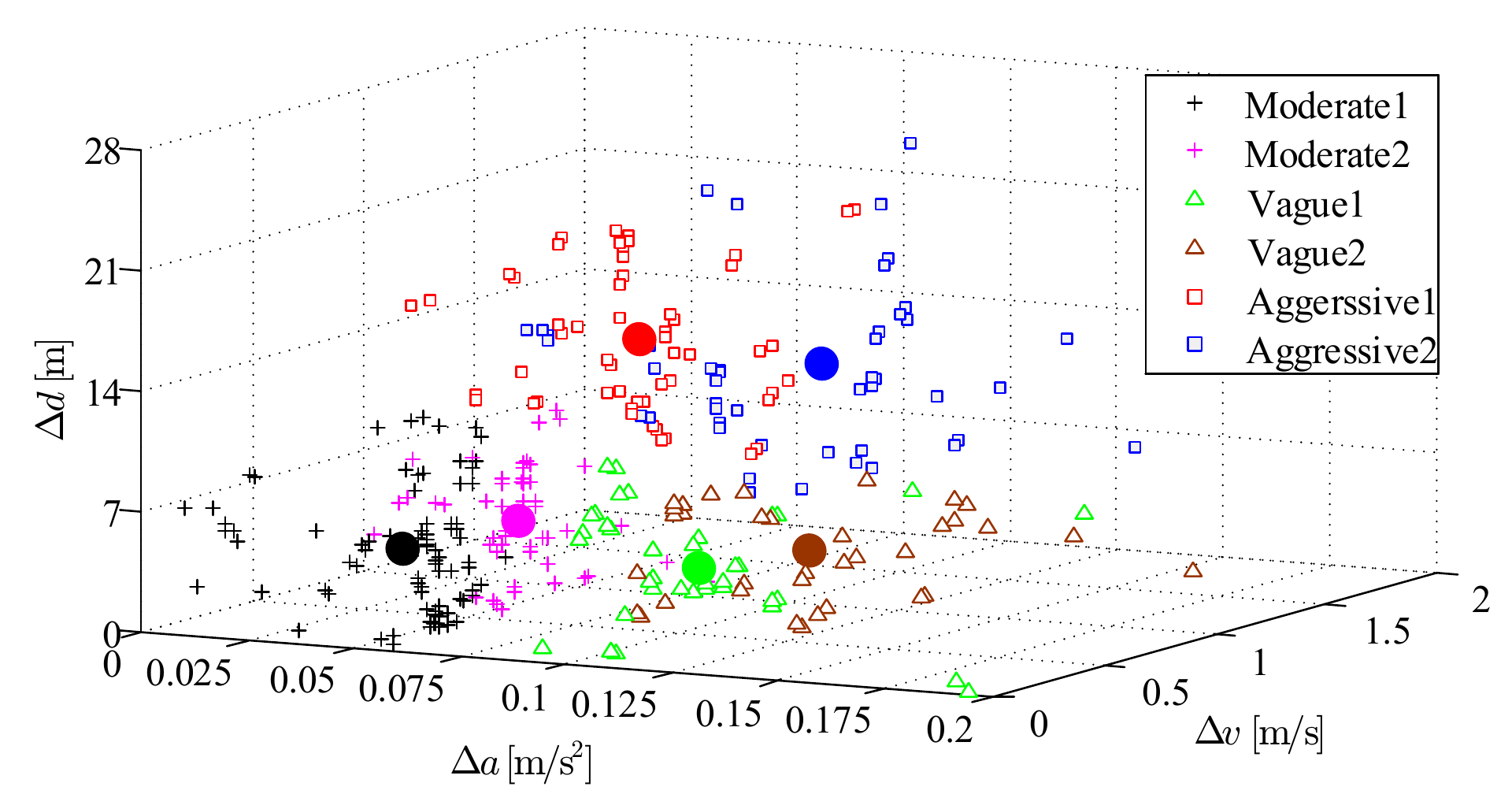}
	\caption{Clustering results of using $k$-MC.}
	\label{Fig.11}
\end{figure}

For the developed $k$MC-KNN recognition method, the $k$-MC is used to partition the raw data in each driving style ($ J $) into $ k $ subsets, thus the raw data is divided into $J\times k$ subsets. For example, the driving data in each diving styles ($J=3$) were divided into $k$ clusters ($ k = 2 $) as shown in Fig. \ref{Fig.11}. Given test data $x^{(*)}$, $k$MC-KNN chooses one cluster from the two clusters based on the similarities between $x^{(*)}$ and the centers in each driving styles as the training samples to reduce the computing cost of KNN.

\subsubsection{Evaluation Metrics}
In order to evaluate recognition performance of $k$MC-KNN, the cross-validation procedure was utilized. For the  $ p $-fold cross-validation, the original data sets were randomly and evenly partitioned into $p$ folds. One single fold was retained as the validation data for testing the model and the remaining $p-1$ folds were used as training data. The cross-validation process was then repeated $p$ times, with each of the $p$ folds used exactly once as the validation data. Totally, $p$ results from all folds then were averaged (or otherwise combined) to get a single estimation result. Here, we randomly partition the original driving data ($N=9936$) into $p=4$ folds ($N_p=2484$) to evaluate the performance of $k$MC-KNN. Then the average accuracy was taken as the final results. The accuracy of the driving-pattern recognizer is computed by
\begin{eqnarray}
\lambda_{j}=\frac{K_{\mathrm{cor},j}}{\sum K_{\mathrm{all},j}}, \ \ j = \mathrm{mod},\mathrm{vag},\mathrm{agg}
\label{eq:15}
\end{eqnarray}
where $\lambda_{j}$ is the accuracy of $j$ driving style. $K_{\mathrm{cor},j}$ is the number of the clustering points that are correctly recognized as $j$ driving style.  $K_{\mathrm{all},j}$ is the number of the clustering points in $j$ driving style.

To show the time-saving performance of $k$MC-KNN, we conducted off-line tests of $k$MC-KNN with different number of clusters clustered by $k$-MC. The traditional KNN  and SVM are chosen as the comparative studies, and the same parameter ($K=\sqrt{N_p}$) is selected for both KNN and $k$MC-KNN, including training data and testing data. The test results using $k$MC-KNN with $k=2,3,4$, and KNN are shown in Table \ref{Tab.4}. $ T $ is the recognition time for $k$MC-KNN and KNN and $ T_0 $ is the recognition time of one data point.
\begin{table*}[t]
	\renewcommand{\arraystretch}{1.3}
	\caption{Comparison Results for KNN and $k$MC-KNN Methods}
	\centering
	\label{Tab.4}	
	\begin{tabular}{c|c|c|c|c|c}
		\hline\hline
		\multicolumn{2}{c|}{\multirow{2}{*}{$K=\sqrt{N_p}$}} & \multirow{2}{*}{KNN}  & \multicolumn{3}{c}{$k$MC-KNN} \\
		\cline{4-6}
		\multicolumn{2}{c|}{} &  & $k=2$ & $k=3$ & $k=4$\\
		\hline
		\multirow{3}{*}{Accuracy} & $\lambda_{mod}$ & 99.06\% ($^{+0.85 \%}_{-0.58 \%}$)& 96.45\% ($^{+2.73 \%}_{-5.74 \%}$) & 96.89\% ($^{+2.11 \%}_{-1.12 \%}$)& 97.11\% ($^{+1.30 \%}_{-1.50 \%}$) \\
		\cline{2-6}
		& $\lambda_{vag}$ & 97.08\% ($^{+2.21\%}_{-2.91\%}$) & 98.26\% ($^{+0.89\%}_{-1.36\%}$) & 95.25\% ($^{+3.75\%}_{-4.30\%}$) & 97.59\% ($^{+0.99 \%}_{-1.93 \%}$) \\
		\cline{2-6}
		& $\lambda_{agg}$ & 94.93\% ($^{+3.60\%}_{-6.25\%}$)  & 91.90\% ($^{+4.51\%}_{-8.27\%}$) & 90.39\% ($^{+5.06\%}_{-9.22\%}$) & 91.16\%  ($^{+5.49\%}_{-6.52\%}$) \\
		\hline
		\multicolumn{2}{c|}{$T$ [s]}& 852.75 ($^{+3.33}_{-1.36}$)  & 233.08 ($^{+10.47}_{-20.92}$) & 131.92 ($^{+15.52}_{-10.60}$) & 91.84 ($^{+10.09}_{-6.98}$) \\
		\hline
		\multicolumn{2}{c|}{$ T_{0} $ [ms]}& 343.30  ($^{+1.34}_{-0.56}$)& 93.83 ($^{+4.25}_{-8.42}$) & 53.11 ($^{+6.25}_{-4.27}$) & 36.97 ($^{+4.06}_{-2.81}$) \\
		\hline\hline
	\end{tabular}
\end{table*}
\begin{figure}[t]
	\centering
	\subfloat[SVM]{\includegraphics[width=0.24\textwidth]{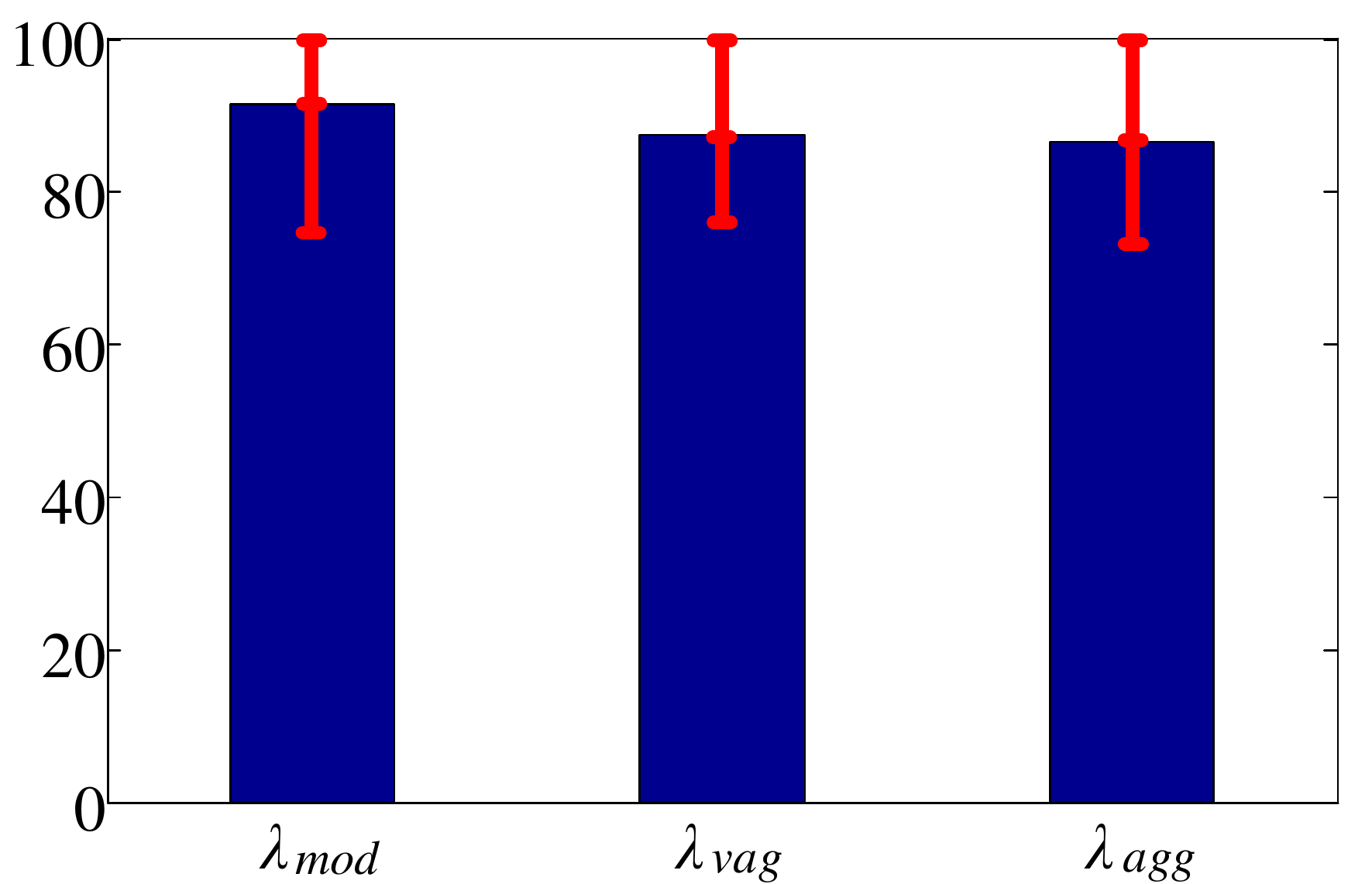}}
	\hfil
	\subfloat[$k=2$ for $k$MC-KNN]{\includegraphics[width=0.24\textwidth]{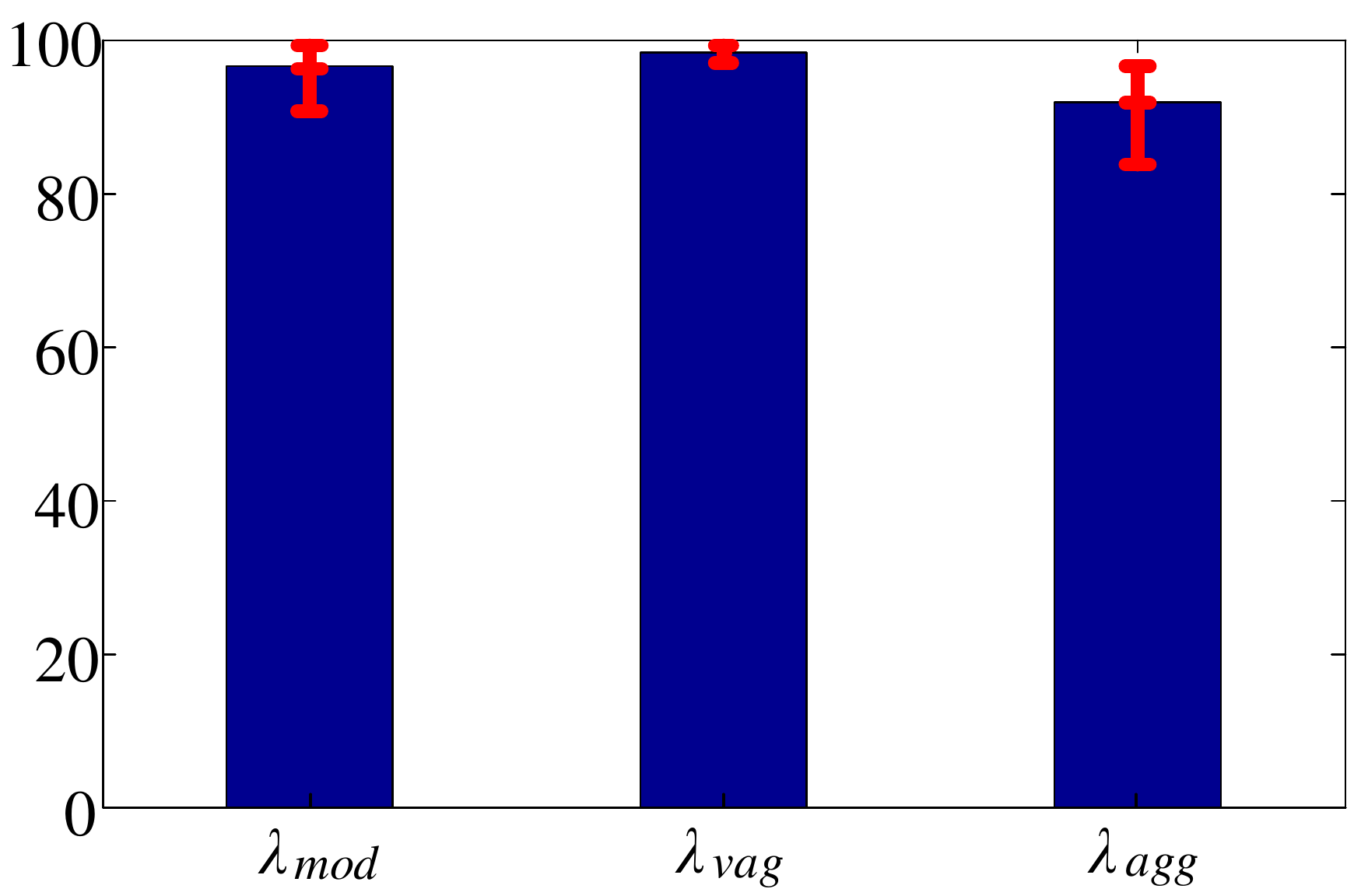}}
	\hfil
	\subfloat[$k=3$ for $k$MC-KNN]{\includegraphics[width=0.23\textwidth]{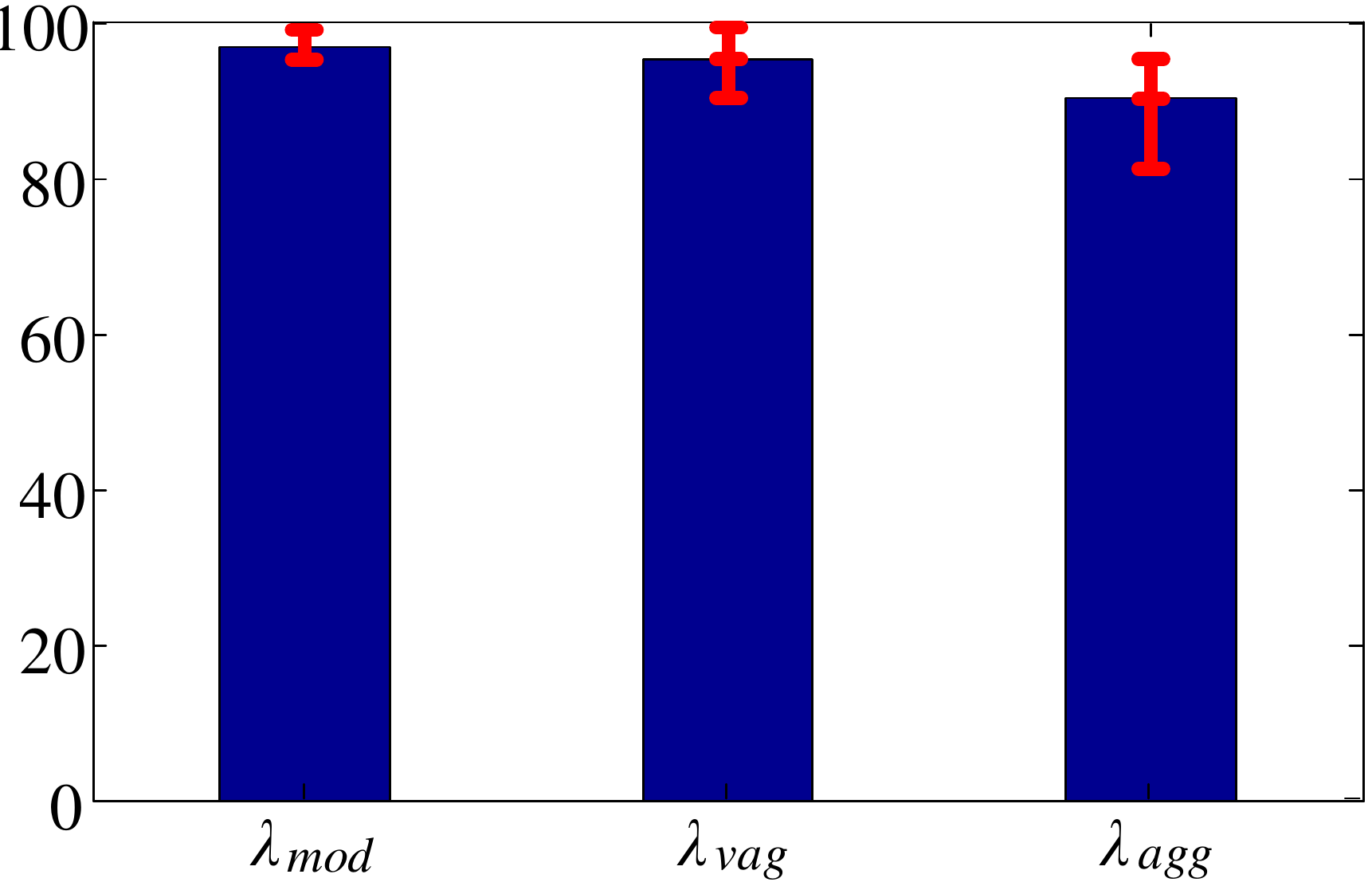}}
	\hfil
	\subfloat[$k=4$ for $k$MC-KNN]{\includegraphics[width=0.24\textwidth]{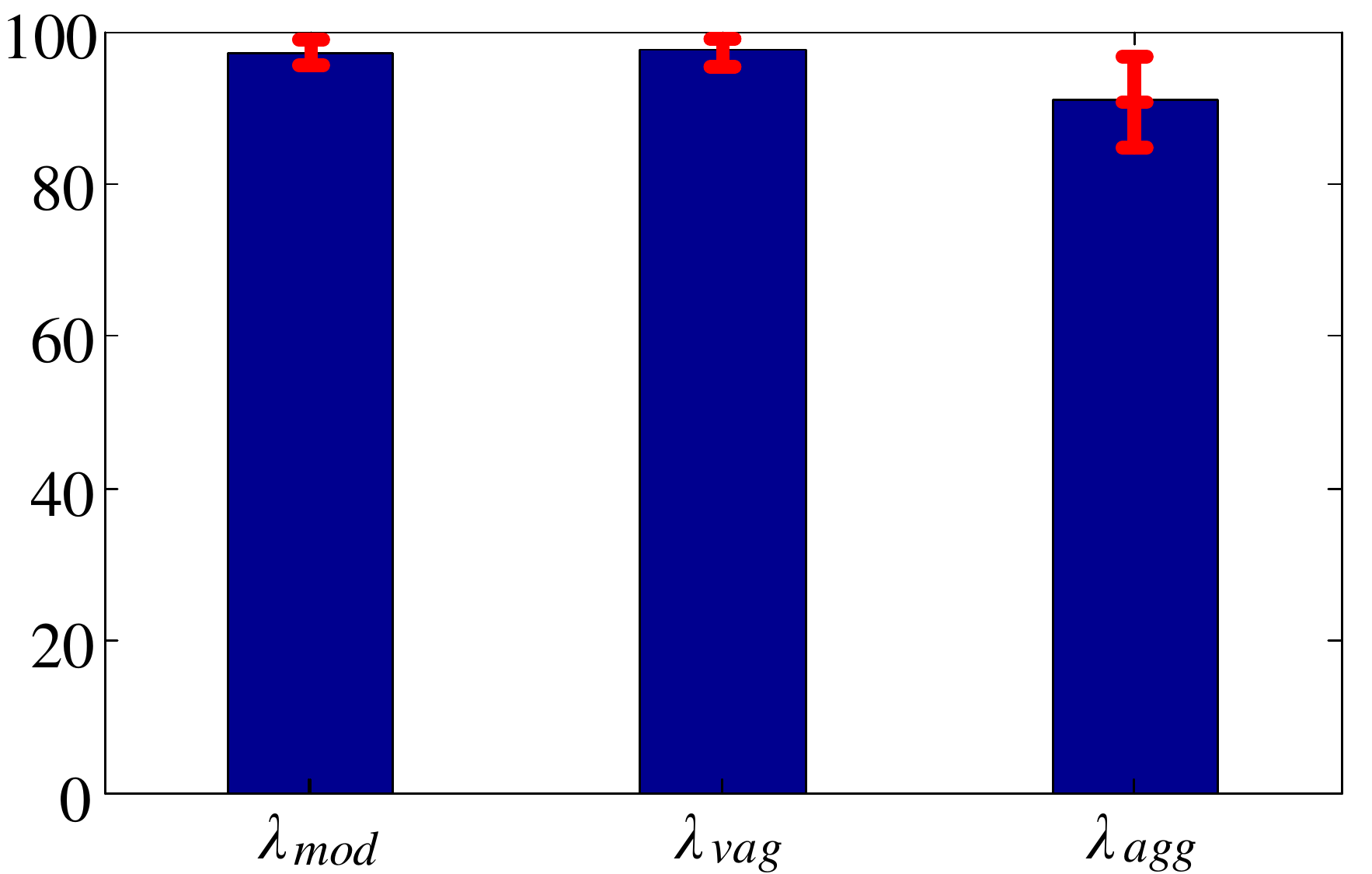}}
	\hfil
	\caption{Comparison recognition accuracy of SVM and our proposed $k$MC-KNN method with $k=2,3,4$.}
	\label{Fig.12}
\end{figure}

\subsubsection{ Result Analysis}
From Table \ref{Tab.4}, it can be seen that the developed $k$MC-KNN outperforms KNN by reducing recognition time significantly over 72.67\%. 
With the increasing value of $k$, the recognition time of $k$MC-KNN gradually decreases since the computation load is decreasing. The accuracy of $k$MC-KNN for vague driving style also outperforms KNN, however the accuracy of both aggressive and moderate driving styles is a slightly lower than that of KNN. The accuracy of $k$MC-KNN fluctuates slightly with increasing value of $k$.

To demonstrate the recognition performance of our proposed method, we also compare it with SVM, as shown in Fig. \ref{Fig.12}. We found that $k$MC-KNN obtains a better performance than SVM. More specifically,  SVM obtains the average recognition accuracy of 87.42\% for vague driving style, while $k$MC-KNN with $k=2$ achieves the accuracy of 98.26\%. Besides, the deviations  (red line) of recognition accuracy also demonstrates that the $k$MC-KNN is more robust than SVM. For example, SVM obtains the average accuracy of vague driving style varying from 76.12\% to 99.86\%, while the $k$MC-KNN with $k=2$ achieves a more stable performance, varying from 96.90\% to 99.15\%.

\section{Conclusions}

This paper developed a $ k $MC-KNN method in order to improve recognition efficiency. An unsupervised clustering method was also proposed based on mathematical morphology in order to reduce efforts of labeling training data and exclude subjective interference from humans. The mathematical morphology-based clustering method can classify drivers' decision-making styles of lane change behavior into three categories with little labeling effort. The experiment results show that our proposed $k$MC-KNN method can shorten recognition time greatly without degrading recognition accuracy. We also found that the developed $k$MC-KNN method outperforms the SVM method in the recognition accuracy and stability. 

\ifCLASSOPTIONcaptionsoff
  \newpage
\fi



%
%
%
\bibliographystyle{IEEEtran}
\bibliography{IEEEexample}
%
%
%

\begin{IEEEbiography}[{\includegraphics[width=1in,height=1.25in,clip,keepaspectratio]{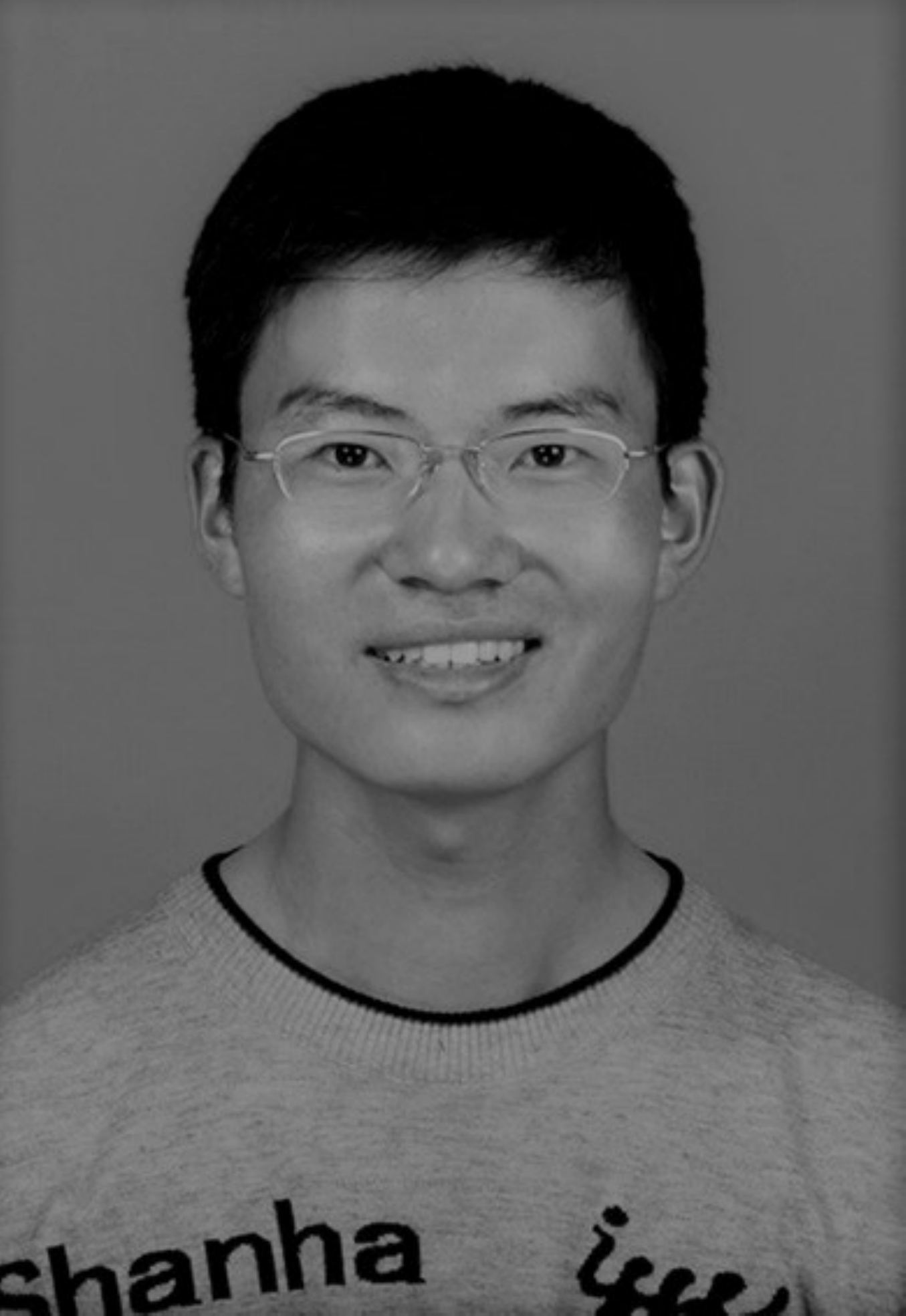}}]{Sen Yang} received the B.S. degree in vehicle engineering from Hubei University of Automotive Technology, Hubei, China, in 2014. He is currently working toward the Ph.D. degree in mechanical engineering with Beijing Institute of Technology (BIT), Beijing, China. His research interests include driver model with artificial intelligence, pattern recognition of human driver characteristics, and human-intelligent vehicle collaboration.
\end{IEEEbiography}
\vspace{-2cm}
\begin{IEEEbiography}[{\includegraphics[width=1in,height=1.25in,clip,keepaspectratio]{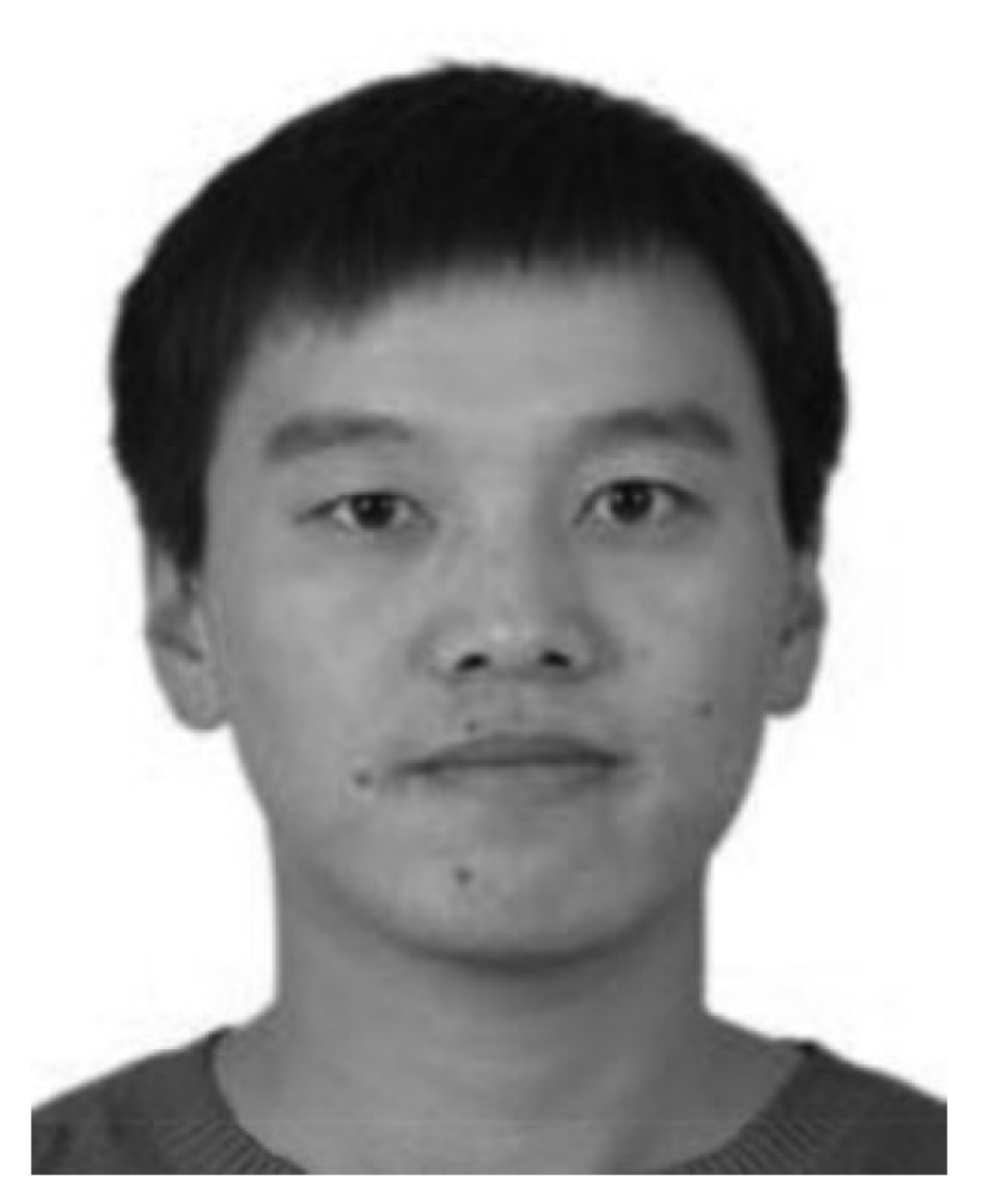}}]{Wenshuo Wang} (S'15-M'18) received his
Ph.D. degree for Mechanical Engineering, Beijing
Institute of Technology (BIT) at June 2018. He is now working as a PostDoc at the Carnegie Mellon University (CMU), Pittsburgh, PA. He also worked as a Research Scholar at the Department of Mechanical Engineering, University of California at Berkeley (UCB) from September 2015 to September 2017 and at the Department of Mechanical Engineering, University of Michigan (UM), Ann Arbor, from September 2017 to July 2018. His research interests include nonparametric Bayesian learning, driver model, human-vehicle
interaction, recognition and application of human driving characteristics.
\end{IEEEbiography}

\begin{IEEEbiography}[{\includegraphics[width=1in,height=1.25in,clip,keepaspectratio]{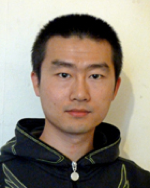}}]{Chao Lu} 
	received the Bachelor’s degree in transport engineering from Beijing Institute of Technology, China, in 2009 and the PhD degree in transport studies from University of Leeds, UK, in 2015. He is currently a lecturer in the School of Mechanical Engineering at Beijing Institute of Technology. His research interests include intelligent transportation and vehicular systems, motorway traffic control, traffic flow modeling, reinforcement learning and its applications. 
\end{IEEEbiography}

\begin{IEEEbiography}[{\includegraphics[width=1in,height=1.25in,clip,keepaspectratio]{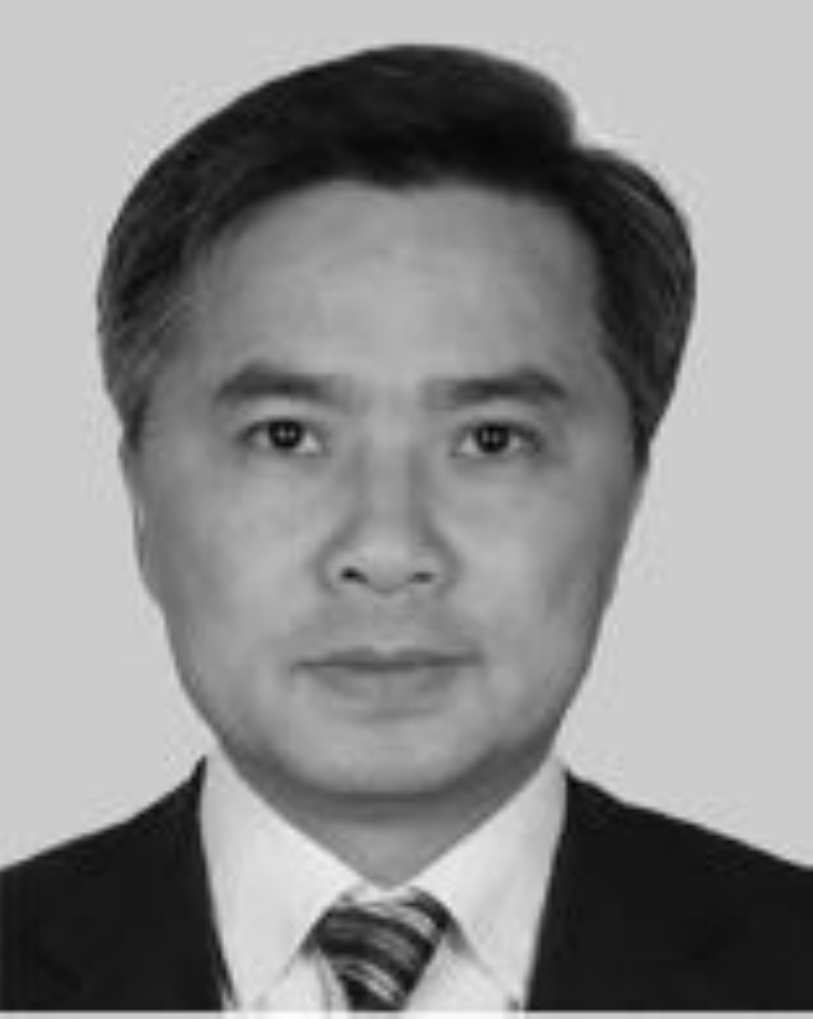}}]{Jianwei Gong} 
	received his B.S. degree from National University of Defense Technology, Changsha, China, in 1992, and Ph.D. degree from Beijing Institute of Technology, Beijing, China, in 2002. He was a visiting scientist of Robotic Mobility Group, Massachusetts Institute of Technology, between 2011 and 2012. He is currently a Professor and Director of the Intelligent Vehicle Research Center, School of Mechanical Engineering, Beijing Institute of Technology. His interests include intelligent vehicle environment perception and understanding,decision making, path/motion planning and control.
\end{IEEEbiography}

\begin{IEEEbiography}[{\includegraphics[width=1in,height=1.25in,clip,keepaspectratio]{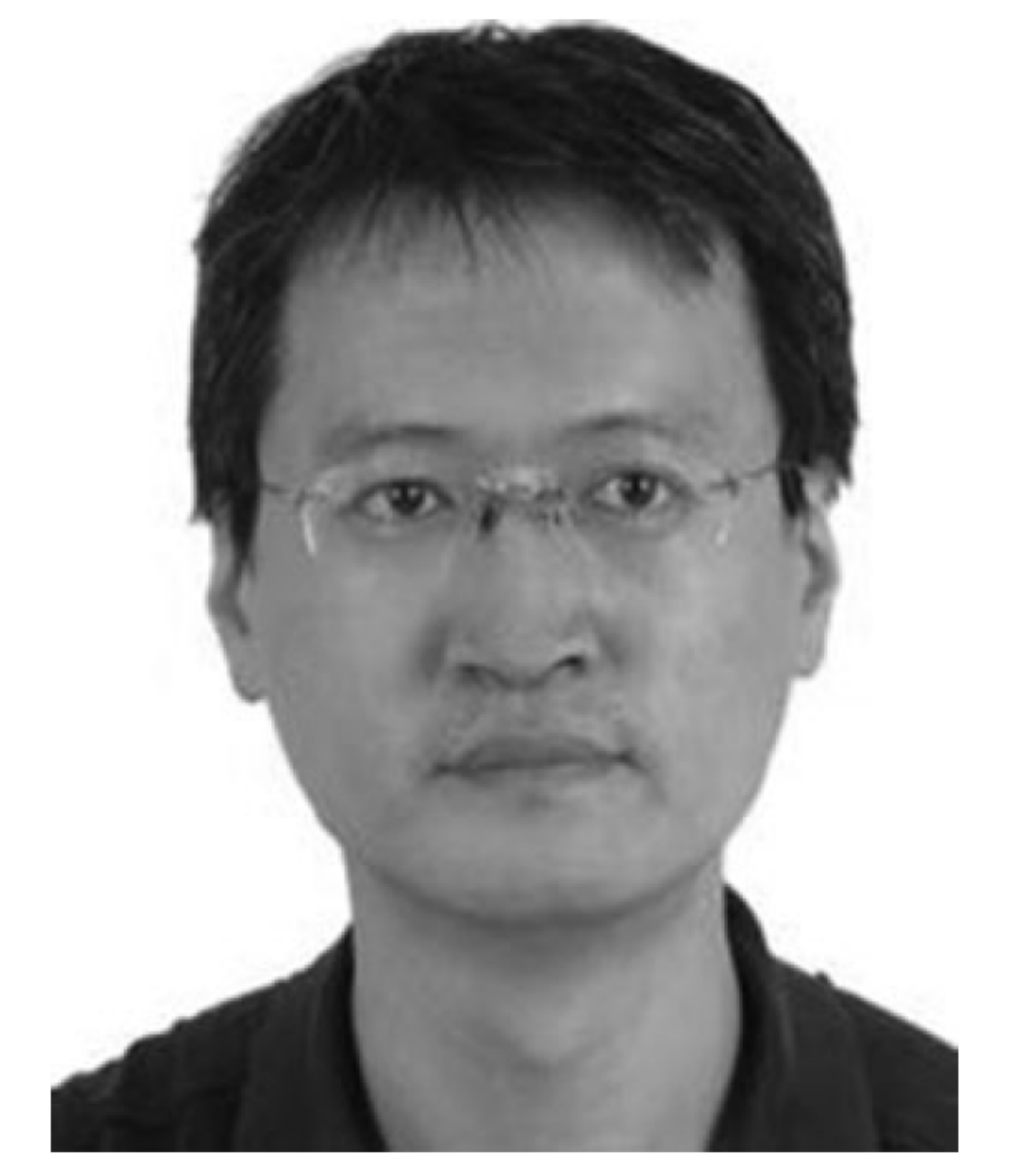}}]{Junqiang Xi} 
	received the B.S. degree in automotive engineering from Harbin Institute of Technology, Harbin, China, in 1995, and the Ph.D. degree in vehicle engineering from Beijing Institute of Technology (BIT), Beijing, China, in 2001.
	
	In 2001, he joined the State Key Laboratory of Vehicle Transmission, BIT. During 2012--2013, he conducted research as an Advanced Research Scholar with the Vehicle Dynamic and Control Laboratory, Ohio State University, USA. He is currently a Professor and Director of the Automotive Research Center, BIT. His research interests include vehicle dynamic and control, powertrain control, mechanics, intelligent transportation systems, and intelligent vehicles.
\end{IEEEbiography}








\end{document}